\documentclass{iopart}
\usepackage{graphicx}
\usepackage{bm}
\usepackage{iopams}
\newcommand{\ba}{\begin{eqnarray}}
\newcommand{\ea}{\end{eqnarray}}
\newcommand{\bse}{\numparts}
\newcommand{\ese}{\endnumparts}
\newcommand{\ACal}{{\cal{A}}}

\newcommand{\DD}{{\cal {D}}}

\newcommand{\W}{{\cal {W}}}
\newcommand{\bbq}{\begin{quote}}
\newcommand{\eeq}{\end{quote}}
\newcommand{\RR}{{}^3{\cal{R}}}

\newcommand{\T}{{}^3{\cal{T}}}

\newcommand{\EE}{{\cal{E}}}
\newcommand{\FF}{{\cal{F}}}

\newcommand{\Vp}{{\cal{V}}_{(p)}}

\newcommand{\HH}{{\cal{H}}}
\newcommand{\KK}{{\cal{K}}}

\newcommand{\Aav}{\langle A\rangle}
 
\newcommand{\hOm}{\Omega_q}
\newcommand{\hOmi}{\Omega_{q0}}
\newcommand{\Dih}{\Delta_0^{(\HH)}}
\newcommand{\Dim}{\Delta_0^{(\rho)}}
\newcommand{\Dik}{\Delta_0^{(\KK)}}
\newcommand{\DiOm}{\Delta_0^{(\Omega)}}
\newcommand{\Daa}{\delta^{(A)}}
\newcommand{\Daai}{\delta_0^{(A)}}
\newcommand{\Diim}{\delta_0^{(\rho)}}
\newcommand{\Diik}{\delta_0^{(\KK)}}
\newcommand{\Diih}{\delta_0^{(\HH)}}
\newcommand{\DiiOm}{\delta_0^{(\Omega)}}

\newcommand{\Da}{\Delta^{(A)}}

\newcommand{\Dh}{\Delta^{(\HH)}}
\newcommand{\DOm}{\Delta^{(\Omega)}}

\newcommand{\Dth}{\Delta^{(\Theta)}}

\newcommand{\Dm}{\Delta^{(\rho)}}

\newcommand{\Dk}{\Delta^{(\KK)}}

\newcommand{\DRR}{\Delta^{(\RR)}}
\newcommand{\dd}{{\rm{d}}}
\newcommand{\tbb}{t_{\textrm{\tiny{bb}}}}

\newcommand{\tcoll}{t_{\textrm{\tiny{coll}}}}

\newcommand{\rr}{\bar r}

\usepackage{pdfsync}

\begin{document}

\title[A novel approach to the dynamics of Szekeres dust models]{A novel approach to the dynamics of Szekeres dust models.}


\author{Roberto A Sussman${}^\dagger$ and Krzysztof Bolejko${}^\ddagger$}
\address{${}^\dagger$ Instituto de Ciencias Nucleares, UNAM, AP 70--543, M\'exico DF, 04510, M\'exico,\\
${}^\ddagger$ Astrophysics Department, University of Oxford, 1 Keble Road, Oxford OX1 3RH, UK}

\ead{sussman@nucleares.unam.mx}
\date{\today}

\begin{abstract}
We obtain an elegant and useful description of the dynamics of Szekeres dust models (in their full generality) by means of ``quasi--local'' scalar variables constructed by suitable integral distributions that can be interpreted as weighed proper volume averages of the local covariant scalars. In terms of these variables, the field equations and basic physical and geometric quantities are formally identical to their corresponding expressions in the spherically symmetric LTB dust models. Since we can map every Szekeres model to a unique LTB model, rigorous results valid for the latter models can be readily generalized to a non--spherical Szekeres geometry. The new variables lead naturally to an initial value  formulation in which all scalars are expressed as scaling laws in terms of their values at an arbitrary initial space slice. These variables also yield a significant simplification of numerical work, since the fluid flow evolution equations become a system of autonomous ordinary differential equations subjected to algebraic constraints containing the information on the deviations from spherical symmetry. 
As an example of how this formalism can be applied, we show that spherical symmetry is stable
against small dipole-like perturbations. This new approach to the dynamics of the Szekeres solutions has an enormous potential for dealing with a wide variety of theoretical issues and for constructing non--spherical models of cosmological inhomogeneities to fit observational data.            
\end{abstract}

\section{Introduction.}

The theory of General Relativity has been successfully employed over the years to describe self--gravitating systems, from the astrophysical up to the cosmological scale. Different phenomena need to be modeled by different solutions of Einstein's field equations. Assuming an idealized description, many self--gravitating systems can be examined by the well known Schwarzschild and Kerr solutions at the astrophysical scale, while the Friedmann--Lema\^itre--Robertson--Walker (FLRW) models provide an adequate idealized framework at the cosmological scale. The extensive use of these simple exact solutions follows from the fact that the applicability of most of the thousands of known exact solutions to model astrophysical or cosmological systems, even as first order idealized approximations, is either impossible, or hard to justify and/or involves accepting unphysical constraints due to the symmetries characterizing these solutions \cite{Step2003}.

A lesser degree of idealization is provided by the well known and often used class of spherically symmetric exact solutions generically known as the  Lema\^\i tre--Tolman--Bondi (LTB) models \cite{Lema1933,Tolm1934,Bond1947}. These models  allow us to examine non--linear effects in self--gravitating systems that have spherical symmetry or can be approximated by spherical configurations by means of exact analytic expressions or, at least, by using numeric but tractable methods. These models have been widely used recently to study the effects of cosmological inhomogeneities within the effort to explain cosmic observations without assuming the existence of a dark energy source (for a review see \cite{BKHC2009,BCK2011}). Since spherical symmetry can be a strong and limiting constraint, even as a first approximation, it is important to look for exact solutions that consider non--spherical generalizations of LTB models, which provide a less idealized description but, at the same time, still allow for either an analytic treatment or a tractable numerical approach. The most general known class of solutions that generalize LTB models in this form are the Szekeres models, which (in general) admit no symmetries (no Killing vectors) and can be reduced to either LTB or FLRW models or the Schwarzschild solutions in the appropriate limits.

The Szekeres solution was found in 1975 by Szekeres \cite{S75} and its first applications was the study of non-spherical collapse \cite{Sz75}. While there is a number of theoretical studies based on Szekeres models \cite{barrow1,barrow2,GoWa1982,HK02,Bsz1,K08,HK08} (see reviews in \cite{PK06,BKHC2009,BCK2011}), only recently these models have gained more interest within the cosmological community as models to
study light propagation \cite{IRGW2008,Bole2009-cmb,BoCe2010,NIT2011,NMMB2011},
or structure formation \cite{Bsz2,IsPe2011,MeBr2011}.
Several new phenomena has been observed within the Szekeres model --
for example the structures can evolve much faster than in the perturbed FLRW or Lema\^itre--Tolman model
\cite{Bsz2,IsPe2011} or that
two rays sent
from the same source at different times to the same observer pass through
different sequences of intermediate matter particles
(as a consequence we should observe a drift of objects positions
in the sky \cite{KrBo2010}).
The Szekeres model was also used to provide a justification of existence
of giant voids \cite{BoSu2011} -- inhomogeneous alternatives to standard cosmological model.

There are two classes of the Szekeres model (see \cite{PK06} for detail):
$\beta' \ne 0$ and $\beta' = 0$ (using the Szekeres notation)
or class I and II (using Goode-Wainwright notation \cite{GoWa1982}).
Class II/$\beta' = 0$ solutions have not received much attention (though
see \cite{IsPe2011,MeBr2011} for recent literature).
Class I/$\beta' \ne 0$ solutions are better known because they are less
idealized and easier to apply as models of inhomogeneities. This class
of solutions has 3 subclasses:
``quasi--spherical'' \cite{HK02,Bsz1}, ``quasi--planar'' \cite{K08}
and ``quasi--hyperbolic'' \cite{HK08} (see \cite{PK06,BKHC2009,BCK2011} for a comprehensive review).
These subclasses are not distinct as within one model
there can be regions of quasi--spherical geometry followed by regions of quasi--hyperbolic geometry \cite{HK08}.

Since Szekeres dust models
\footnote{Szekeres models with uniform pressure where first examined
by Szafron \cite{Szafron}, see also \cite{kras97}.
For the use of q--scalars with nonzero pressure in the spherically symmetric case
see  \cite{suss09}.}
 are characterized by vanishing vorticity, 4--acceleration and  magnetic Weyl tensor, they belong to the class of so--called ``silent universes'', which are appropriate approximations to a general spacetime in the long wave limit \cite{lw1,lw2,lw3}.  Silent universes (in general) suffer from various problematic features, such as: linear instability and non--integrability of their spatial constraints in a fluid flow formalism \cite{silent1,silent2}. 
However, such problems do not arise for the case of Szekeres models
and their higher symmetry subcases. As commented in \cite{silent1,silent2}, there is a strong conjecture that Szekeres models may be the only spatially inhomogeneous self--consistent silent universes.

We aim in this paper to present a novel approach to the study of the
Szekeres models of class I ($\beta' \ne 0$) that is based on new coordinate independent scalar variables (``quasi-local'' variables or ``q--scalars'').  These variables have been previously introduced in the study of  spherically symmetric LTB models \cite{suss2009, sussIU, suss2010a}, looking at important theoretical issues, such as exact non--linear perturbations \cite{suss2009}, averaging inhomogeneities \cite{sussIU,LTBave1,LTBave2,sussBR1,sussBR2}, radial asymptotics \cite{RadAs}, evolution of radial profiles \cite{RadProfs}, as well as a dynamical systems analysis \cite{ds1,ds2}. By extending the use of these variables to class I Szekeres  models we can generalize these studies to non--spherical geometries.

Since the q--scalars are defined in terms of suitable integral distributions of the local covariant fluid flow scalars (density, Hubble flow, spatial curvature), they become weighed averages of these local covariant scalars if defined as functionals (instead of functions). By expressing the local scalars as fluctuations of the q--scalars, these fluctuations together with the q--scalars and algebraic constraints linking them provide a scalar representation that completely determines that dynamics of the models, either in terms of analytic scaling laws or by means of autonomous evolution equations suitable for numeric work.

It is a well known fact that Einstein's field equations reduce for all silent universes to a system of six fluid flow (or ``1+3'') evolution equations that contain no spatial derivatives, and thus can be treated (formally) as a system of six ODE's \cite{baro,MPS1,MPS2,bruni1,bruni2,croud}. If we use the standard fluid flow variables (as in the sources cited above) the spacelike constraints become a set of complicated  non--linear PDE's on the spacelike coordinates. Since these PDE's must be satisfied al all times, they necessarily constrain initial data and make it harder to conduct the numeric work of integrating the evolution equations. However, if we use the q--scalars to construct a set of evolution fluid flow equations, not only we can handle these equations  as a system of ODE's, but the PDE's that provide the spacelike constraints in the standard variables reduce to algebraic constraints, which are formally identical to those of LTB models, with the deviation from spherical symmetry entering through initial conditions. Hence, the new variables lead to a simplified dynamics for the Szekeres models and also allow for a better understanding of effects of their deviation from spherical symmetry. Evidently, dealing with algebraic constraints that are supplied by means of initial data represents a valuable advantage over the traditional fluid flow variables used in \cite{baro,MPS1,MPS2,bruni1,bruni2,croud}. Also, the q--scalars may provide important information on key theoretical aspects, as well as illuminate the connection between the Szekeres and LTB models and with linear perturbations on a FLRW background.

The section by section plan of the paper is as follows. We introduce in section \ref{SzekLTBlike} the Szekeres models given in a parametrization that expresses their main quantities as formally identical to the corresponding quantities in LTB models. Field equations, covariant fluid flow scalars and their evolution equations are given in terms of this parametrization. A new set of quasi--local (``q--scalar'') variables is introduced in section \ref{QLV}, together with fluctuations defined by comparing the local fluid flow variables with their associated q--scalars. We discuss briefly in section \ref{QLVprops} the relation between the q--scalars and averages, as well as with the decomposition of Szekeres scalars into monopole and dipole--like terms. Fluid flow evolution equations for the q--scalars and their fluctuations are obtained in various representations in section \ref{QLVeveqs}, showing how the deviation from  spherical/planar/hyperbolic symmetry can be fully accounted for as initial conditions are specified. Initial conditions for these evolution equations are discussed in section \ref{Initcond}. We introduce an initial value formulation in section \ref{Initval}, leading to analytic solutions of the evolution equations in terms of the scale factors and scaling laws for the q--scalars and their fluctuations. In section \ref{Regconds} we examine the regularity conditions to avoid shell crossings, which in the framework of this formulation can be stated as restrictions on the initial value functions. We show in section \ref{CompLTB} how we can associate to each LTB model  a 3--parameter class of Szekeres models by a simple transofrmation in the space of initial conditions. Each Szekeres model in this class is characterized by the same q--scalars as the LTB model, and its fluctuations can be obtained from the LTB fluctuations by a simple algebraic relation. In section \ref{dipole} we apply the initial value formalism to examine the stability of the deviation from spherical conditions by looking at the evolution of the dipole contribution in quasi--spherical Szekeres models close to spherical symmetry. In section \ref{Final} we summarize our results and provide a discussion of potential applications. We have included  four appendices that complement the material covered in the article: scalar averaging in Szekeres models (\ref{averaging}), covariant expressions for the q--scalars and their fluctuations (\ref{covariant}), the proof that functions of q--scalars are also q--scalars (\ref{qfunctions}) and models with spherical or wormhole topologies (\ref{wormholes}).

\section{Szekeres models in the ``LTB--like'' parametrization.\label{SzekLTBlike}}

We begin with the spherically symmetric Lema\^\i tre--Tolman--Bondi (LTB) metric in its standard representation:
\begin{equation}
\dd s^2 = -\dd t^2 + \frac{R'{}^2}{1-K}\dd r^2 + R^2\left(\dd\theta^2 +\sin^2\theta\dd\phi^2\right),\label{ltbmetric}
\end{equation}
where $R=R(t,r),\,R'=\partial R/\partial r$ and $K=K(r)$. The corresponding field equations for a dust tensor and a comoving 4--velocity $u^a=\delta^a_0$ take the well known form
\begin{equation}  \dot R^2 = \frac{2M}{R}-K,\qquad 2M' = 8\pi \rho\,R^2 R',\label{efeltb} \end{equation}
where $M=M(r)$ is the quasi--local or Misner--Sharp mass function and $\dot R =u^a \nabla_a R=\partial R/\partial t$. It is straightforward to write the metric of the Szekeres model in a form that is similar to (\ref{ltbmetric}):  
\begin{equation}\dd s^2 = -\dd t^2 + \frac{\EE^2\,Y'{}^2}{\epsilon-K}\dd r^2 + Y^2\left[\dd x^2 + \dd y^2\right],\label{szmetric}\end{equation}
where $Y=Y(t,r,x,y)$ and $\EE=\EE(r,x,y)$ are given by
\ba Y = \frac{R}{\EE},\label{Ydef}\\
 \EE = \frac{S}{2}\,\left[\epsilon+\left(\frac{x-P}{S}\right)^2+\left(\frac{x-Q}{S}\right)^2\right],\label{Edef}\ea
with $S(r),\,P(r),\,Q(r)$ arbitrary functions, and $\epsilon = 0, \pm 1$. 
The cases $\epsilon = 1$, $0$, and $-1$,
respectively, correspond to quasi-spherical, quasi-planar, and
quasi-hyperbolic models.

The field equations take the same LTB--like form
\ba \dot Y^2 = \frac{2\tilde M}{Y}-\tilde K,\label{efe1}\\
 2\tilde M' = 8\pi\rho \,Y^2 Y',\label{efe2}\ea
where 
\begin{equation}\tilde M = \frac{M}{\EE^3},\qquad  \tilde K = \frac{K}{\EE^2}.\label{tildeMK}\end{equation}
Relevant covariant quantities take also the LTB--like form: the expansion scalar $\Theta = \nabla_a u^a$ and the Ricci scalar $\RR$ of hypersurfaces $t=$ const (orthogonal to $u^a$)  
\ba \Theta = \frac{2\dot Y}{Y}+\frac{\dot Y'}{Y'},\label{Theta1}\\
\RR = \frac{2(\tilde K Y)'}{Y^2Y'},\label{RR1}\ea
together with the shear and electric Weyl tensors
\ba \fl \sigma_{ab} = \tilde \nabla_{(a}u_{b)}-\frac{\Theta}{3}h_{ab}=\Sigma\, \Xi_{ab},\qquad \Sigma = \Xi_{ab}\sigma^{ab}=-\frac{1}{3}\left(\frac{\dot Y'}{Y'}-\frac{\dot Y}{Y}\right),\label{Sigma1}\\
\fl  W^{ab} = u_cu_d C^{acbd}= \W\,\Xi^{ab},\qquad \W = \Xi_{ab}W^{ab}=-\frac{\tilde M}{Y^3}+\frac{4\pi}{3}\,\rho,\label{W1}\ea
where $C^{acbd}$ is the Weyl tensor, $h_{ab}=g_{ab}+u_au_b$, \,\,$\tilde \nabla_a = h_a^b\nabla_b$ and $\Xi_{ab}=h_{ab}-3\eta_a\eta_b$, with $\eta_a=\sqrt{h_{rr}}\delta_a^r$. The quantities (\ref{Theta1})--(\ref{W1}) become identical to their LTB forms by replacing $Y,\,\tilde K,\,\tilde M$ for $R,\,K,\,M$. 

The standard procedure to deal with the Szekeres model is to solve the Friedmann--like equation (\ref{efe1}) by means of the following quadrature, which is equivalent to the LTB equation for $\dot R$ in (\ref{efeltb}):
\begin{equation}
t-\tbb(r) = \int_{u=0}^{u=R}{\frac{\dd u}{\sqrt{2M/u -K}}},\label{quadrature}
\end{equation}
where we eliminated $Y, \tilde{M},\tilde{K}$ in (\ref{efe1}) in terms of  $R, M, K$ by means of (\ref{Ydef}) and (\ref{tildeMK}).
Above $\tbb(r)$ is another arbitrary function, the time locus of the big bang singularity, which adds to the available $r$--dependent free functions $S,\,Q,\,P,\,K,\,M$. Considering that the metric (\ref{szmetric}) allows for an arbitrary rescaling $r=r(\rr)$, so that any one of these functions can be eliminated by a suitable choice of the $r$ coordinate, then the solution of (\ref{quadrature}) for a given choice of free functions determines a specific model in which the density follows from (\ref{efe2}) and the remaining relevant quantities from (\ref{Theta1})--(\ref{W1}).

Since the scalars 
$\{\rho,\,\Theta,\,\RR,\,\Sigma,\,\W\}$
 completely characterize the Szekeres models (considering (\ref{Sigma1}) and (\ref{W1})), an alternative to dealing with the dynamics of these models through the quadrature (\ref{quadrature}) is furnished by the ``1+3'' or fluid--flow approach \cite{1plus3}, leading to the following scalar evolution equations:~\cite{1plus3}
\bse\ba \dot \rho = -\rho\,\Theta,\label{FFev1}\\
\dot \Theta =-\frac{\Theta^2}{3}-\frac{\kappa}{2}\,\rho-6\,\Sigma^2,\label{FFev2}\\
\dot \Sigma = -\frac{2}{3}\Theta\,\Sigma-\Sigma^2+\W,\label{FFev3}\\
\dot \W = -\Theta\, \W -\frac{\kappa}{2}\rho\,\Sigma+3\Sigma\,\W,\label{FFev4}\ea\ese
together with the ``Hamiltonian'' constraint
\begin{equation}
\frac{\Theta^2}{9} = \frac{\kappa\,\rho}{3}-\frac{\RR}{6}+\Sigma^2.\label{FFham}
\end{equation}
and the spacelike constraints:
\begin{equation} \tilde\nabla_b\sigma^b_a-\frac{2}{3}h_a^b\Theta_{,b}=0,\qquad
\tilde\nabla_b W^b_a-\frac{\kappa}{3}h_a^b\rho_{,b}=0, \label{FFcon}\end{equation}
which take the following form (component by component):
\bse\ba \Sigma' + \frac{\Theta'}{3}+\frac{3Y'}{Y}\Sigma=0,\label{FFcon1}\\
\W' + \frac{\kappa}{6}\rho'+\frac{3Y'}{Y}\W=0,\label{FFcon2}\\
\frac{(\EE'/\EE)_{,x}}{Y'/Y}-\frac{\Sigma_{,x}}{3\Sigma}=0,\qquad \frac{(\EE'/\EE)_{,y}}{Y'/Y}-\frac{\Sigma_{,y}}{3\Sigma}=0,\label{FFcon3}\\
\frac{(\EE'/\EE)_{,x}}{Y'/Y}-\frac{\W_{,x}}{3\W}=0,\qquad \frac{(\EE'/\EE)_{,y}}{Y'/Y}-\frac{\W_{,y}}{3\W}=0,\label{FFcon4}\ea\ese
where ${}_{,x}$ and ${}_{,y}$ respectively denote $\partial/\partial x$ and $\partial/\partial y$. 

Notice that the fluid flow system presented above is identical to that in the ``1+3'' fluid flow formalism \cite{1plus3} applied to LTB models \cite{zibin,dunsbyetal}, save for the constraints (\ref{FFcon3}) and (\ref{FFcon4}) which contain the input from the ``non--spherical'' degrees of freedom through dependence on $(x,y)$. Evidently, these constraints distinguish a Szekeres and an LTB model. Also, as a consequence of (\ref{FFcon3}) and (\ref{FFcon4}), the quotient $\W/\Sigma$ only depends on $(t,r)$, though $\W$ and $\Sigma$ each depends also on $(x,y)$. 

\section{Quasi--local scalar variables.}\label{QLV}
The hypersurfaces $\T[t]$ marked by $t=$ constant, whose Ricci scalar is given by (\ref{RR1}), provide a natural time slicing for the Szekeres models. The $\T[t]$ in Szekeres models, in general, admit a foliation in terms of compact simply connected domains $\DD\subset \T[t]$ parametrized by the coordinates $(r,x,y)$ 
(see \ref{averaging} \footnote{The worldline $r=0$ cannot be contained in any domain $\DD$ of quasi--hyperbolic models ($\epsilon =-1$). For quasi--planar and quasi--spherical models ($\epsilon =1$) this worldline can be characterized as a special location where  $R=Y=\Sigma=\W=0$ hold for all $t$. It is not a symmetry center in quasi--spherical models, since the latter are not spherically symmetric. }). 
For every scalar function $A$ defined in $\DD$ in an arbitrary $\T[t]$ we can generate the ``quasi--local''  
scalar function (or ``q--scalar'') $A_q:\DD\to \mathbb{R}$ dual to the local scalar $A$ by means of the following correspondence rule: 
\begin{equation}
A_q = \frac{\int_\DD{A\FF\dd\Vp}}{\int_\DD{\FF\dd\Vp}}=\frac{\int_r {\int_x {\int_y {A \ \EE Y^2 Y' \dd r \dd x \dd y} } } }{\int_r {\int_x {\int_y {\EE Y^2 Y' \dd r \dd x \dd y} } } }, \label{qdef}
\end{equation}
where $\FF =\sqrt{\epsilon-K}$ and $\dd\Vp$ is the proper volume element $\dd\Vp=\FF^{-1} \EE Y^2Y'\dd r\dd x\dd y$ of the time slices $\T[t]$. 
\footnote{The term ``quasi--local'' follows from the integral definition of the quasi--local Misner--Sharp mass--energy function in spherical symmetry \cite{suss2009}. The relation between (\ref{qdef}) and the average of $A$ with weight factor $\FF$ over the domain $\DD$ is discussed in section 4.1 (see also \cite{sussIU,sussBR1,sussBR2}).}

Considering that (\ref{Theta1}) implies $\Theta = [\ln (Y^2Y')]\,\dot{}$, together with the forms of $\rho$ and $\RR$ in (\ref{efe2}) and (\ref{RR1}), we obtain by applying (\ref{qdef}) to $A=\rho,\,\Theta,\,\RR$ the following closed analytic forms:
\ba \frac{8\pi}{3}\rho_q = \frac{2\tilde M}{Y^3}=\frac{2M}{R^3},\label{rhoq}\\
\frac{\Theta_q}{3} =\frac{\dot Y}{Y} = \frac{\dot R}{R},\label{Thetaq}\\
\frac{\RR_q}{6} = \frac{\tilde K}{Y^2} = \frac{K}{R^2},\label{RRq}
\ea
so that the q--scalars $\rho_q,\,\Theta_q,\,\RR_q$ only depend on $(t,r)$, even if their ``local'' counterparts $\rho,\,\Theta,\,\RR$ depend (in general) on all four coordinates. 
\footnote{While the q--scalars $\rho_q,\,\Theta_q,\,\RR_q$ are coordinate independent quantities (see \ref{covariant}), the function $\tilde M$  given in (\ref{efe2}) is not an invariant scalar of the Szekeres models (as its equivalent $M$ is for spherically symmetric spacetimes).}
The relation between (\ref{rhoq})--(\ref{RRq}) and the weighed averages $\langle\rho\rangle_q,\,\langle\Theta\rangle_q,\,\langle\RR\rangle_q$ obtained from treating (\ref{qdef}) as a functional is discussed in \ref{averaging} (see section 4.1 and \cite{sussIU,sussBR1,sussBR2}). 

By differentiating (\ref{rhoq}), (\ref{Thetaq}) and (\ref{RRq}) with respect to $r$ and applying (\ref{efe2}), (\ref{Theta1}) and (\ref{RRq}) we obtain
\begin{equation}
\fl \rho'_q = \frac{3Y'}{Y}\left(\rho-\rho_q\right),\qquad \Theta'_q = \frac{3Y'}{Y}\left(\Theta-\Theta_q\right),\qquad \RR'_q = \frac{3Y'}{Y}\left(\RR-\RR_q\right),\label{Aqr}
\end{equation}
which implies, by means of (\ref{Theta1}), (\ref{Sigma1}) and (\ref{W1}), that the scalars associated with the shear and electric Weyl tensors are expressible as fluctuations of $\Theta$ and $\rho$ with respect to their q--scalar duals $\Theta_q$ and $\rho_q$:
\begin{equation}
\Sigma = -\frac{1}{3}(\Theta-\Theta_q),\qquad \W = -\frac{4\pi}{3}(\rho-\rho_q).\label{SigW}
\end{equation}
In order to connect local and q--scalars, we define the relative fluctuations of $\rho,\,\Theta,\,\RR$ as
\begin{equation}
\Da \equiv  \frac{A-A_q}{A_q}= \frac{A'_q/A_q}{3Y'/Y},\qquad A = \rho,\,\Theta,\,\RR,\label{Dadef}
\end{equation}
where we used the derivation rule (\ref{Aqr}). Since $\rho_q,\,\Theta_q,\,\RR_q$ are independent of $(x,y)$, we have from (\ref{Dadef}) for $A = \rho,\,\Theta,\,\RR$:
\bse\ba
\Da_{,x} = -\Da\,\frac{(Y'/Y)_{,x}}{Y'/Y}=\Da\,\frac{(\EE'/\EE)_{,x}}{Y'/Y},\label{Dax}\\
\Da_{,y} = -\Da\,\frac{(Y'/Y)_{,y}}{Y'/Y}=\Da\,\frac{(\EE'/\EE)_{,y}}{Y'/Y},\label{Day}
\ea\ese
where we used $Y'/Y=R'/R-\EE'/\EE$ from (\ref{Ydef}). As shown in \ref{covariant}, the fluctuations (\ref{Dadef}) are coordinate independent quantities related to curvature invariants. 

\section{Some appealing properties of the q--scalars.}\label{QLVprops}

\subsection{The q--scalars as averages.}\label{QLVave} 

If we use the correspondence rule (\ref{qdef}) to define functionals (not functions) then we can interpret the q--scalars as proper volume average distributions $\Aav_q$ (see \ref{averaging}) on a domain $\DD$ with weight factor
\begin{equation} \FF=\sqrt{\epsilon-K}=\left[\dot Y^2+\left(\epsilon-\frac{2\tilde M}{Y}\right)\right]^{1/2},\label{FF}\end{equation}
which in spherical symmetry is a scalar invariant that reduces to the ``$\gamma$'' factor in the Special Relativity limit 
and to total (or ``binding'') energy in the Newtonian limit \cite{hayward}. The difference between the q--scalar and the q--average is subtle: for any given domain $\DD$ the q--average associates the number $\Aav_{q\DD}$ to the whole domain $\DD$, while the q--scalars are pointwise local functions $\DD\to \mathbb{R}$, hence both are equal only at the boundary  $\partial\DD$ of each domain (which is marked by $r$ constant). As a consequence, both satisfy the same local derivation rules at the boundary of an arbitrary $\DD$, but behave differently under integration over the whole $\DD$. An elaborate discussion on the relation between q--scalars and q--averages is given in \cite{sussIU,sussBR1,sussBR2} for the spherically symmetric case (though it applies to the Szekeres case). 

As shown in \cite{Bole2009-vav}, the standard proper volume average (scale factor = 1) of a local scalar $A$ in all Szekeres models is independent of $(x,y)$. We show explicitly in \ref{averaging} that this is also the case for the q--scalars.  Unless explicitly stated otherwise, we will consider throughout the article the $A_q$ obtained from (\ref{qdef}) as new local functions whose correspondence rule is the same as that of the weighed average functional $\Aav_q$.

\subsection{Decomposition in terms of a monopole and a dipole.}\label{QLVsum}

It is important to remark that relevant physical and geometric scalar quantities in Szekeres models, such as density, expansion rate, etc, can be decomposed as a sum of a pure ``radial'' part and a ``non--radial remainder''\footnote{In quasi--spherical case this ``remainder'' has the structure of a dipole,
while in the quasi-hyperbolic case it is a pseudo-spherical equivalent of a dipole.}, {\it i.e.}
\begin{equation} A = A_{\textrm{\tiny{rad}}} + A_{\textrm{\tiny{non-rad}}},\qquad A_{\textrm{\tiny{rad}}} \equiv A|_{\EE'=0},\label{mondip}\end{equation}
where $A_{\textrm{\tiny{rad}}}$ only depends on $t$ and $r$. As shown in \cite{Bole2009-vav}, in computing the standard proper volume average of $A$ (which is (\ref{qdef}) with $\FF=1$) for quasi--spherical models ($\epsilon=1$)  the dipole cancels out, so that $\Aav = \langle A_{\textrm{\tiny{rad}}}\rangle$. We prove in \ref{averaging} that 
this also holds for the q--average with weight factor $\FF\ne 1$ with the ``angular'' part canceling out also in the 
quasi-hyperbolic and quasi-plane cases, i.e. we have for every domain $\DD[r]$ bounded by a surface $r=$ constant:
\begin{equation}
A_q(r) = \frac{\int_r {\int_x {\int_y {A \ \EE Y^2 Y'drdxdy} } } }{\int_r {\int_x {\int_y {\EE Y^2 Y'drdxdy} } } } = \frac{ \int_r  {A_{\textrm{\tiny{rad}}} R^2 R' dr} }{{\int_r R^2 R'dr} }=\Aav_{q\DD[r]},
\label{qdef2}
\end{equation}
where we are assuming that these integrals are bounded for a given domain contained in the slices $\T[t]$ in models that are not quasi--spherical. As a consequence, we can think of q--scalars as providing at every domain $\DD[r]$ the q--average of the monopole ``radial'' term in the decomposition (\ref{mondip}).

\section{Evolution equations for the quasi--local variables.}\label{QLVeveqs}

Considering (\ref{SigW}) and (\ref{Dadef}), the five scalars $\{\rho,\,\Theta,\,\Sigma,\,\W,\,\RR\}$ that characterize the fluid flow dynamics of the Szekeres models are expressible in terms of the scalar representation $\{A_q,\,\Da\}$ for $A = \rho,\,\Theta,\,\RR$:  
\bse\ba \fl \rho = \rho_q(1+\Dm),\qquad \HH = \HH_q(1+\Dh),\qquad \KK = \KK_q(1+\Delta^{(\KK)}),\label{locvars1}\\
\fl \Sigma = -\HH_q\Dh,\qquad \W = -\frac{4\pi}{3}\rho_q\Dm,\label{locvars2}
 \ea\ese  
where we are using (and will use henceforth) the notation
\begin{equation} \HH \equiv \frac{\Theta}{3},\qquad \KK \equiv \frac{\RR}{6}.\label{notation}\end{equation}
so that $\Dth = \Dh,\,\DRR = \Dk$. 

\subsection{Evolution equations and constraints.}\label{QLVevcons}

Eliminating the local scalars $\{\rho,\,\HH,\,\Sigma,\,\W\}$ in terms of $\{\rho_q,\,\HH_q,\,\Dm,\,\Dh\}$ by means of (\ref{locvars1})--(\ref{locvars2}) and substituting into the 1+3 system (\ref{FFev1})--(\ref{FFev4}) yields the following system of autonomous evolution equations
\bse\ba \dot \rho_q &=& -3 \rho_q\HH_q,\label{FFq1}\\
\dot \HH_q &=& -\HH_q^2-\frac{4\pi}{3}\rho_q, \label{FFq2}\\
\dot\Delta^{(\rho)} &=& -3(1+\Dm)\,\HH_q\Dh,\label{FFq3}\\
\dot\Delta^{(\HH)} &=& -(1+3\Dh)\,\HH_q\Dh+\frac{4\pi\rho_q}{3\HH_q}(\Dh-\Dm),\label{FFq4}\ea\ese
while substitution of (\ref{locvars1})--(\ref{locvars2}) in (\ref{FFham}) and (\ref{FFcon}) yields the algebraic constraints 
\ba \HH_q^2 = \frac{8\pi}{3}\rho_q - \KK_q,\label{FFqham}\\
2\Dh = \hOm\,\Dm + (1-\hOm)\,\Dk,\label{FFqcon}\ea
where we have introduced the following q--scalar analogous to a FLRW Omega factor 
\begin{equation}\hOm \equiv \frac{8\pi\rho_q}{3\HH_q^2},\qquad \hOm -1 = \frac{\KK_q}{\HH_q^2}.\label{Omdef}\end{equation}
whose corresponding fluctuation is (from (\ref{Dadef})) 
\begin{equation} \DOm = \Dm -2\Dh =(1-\hOm)(\Dm-\Dk),\label{DOmdef}\end{equation}
and its local dual is $\Omega=\hOm(1+\DOm)$. As we show in \ref{qfunctions}, all scalars expressible as functions of q--scalars are also q--scalars and comply with (\ref{qdef}) and (\ref{Dadef}).

We remark that the constraints (\ref{FFcon1})--(\ref{FFcon2}), associated with the ``radial'' derivatives of $\Sigma$ and $\W$, reduce to the first two identities in (\ref{Aqr}), while the constraints (\ref{FFcon3})--(\ref{FFcon4}), associated with the ``non--radial'' derivatives of $\Sigma$ and $\W$, reduce to the identities (\ref{Dax})--(\ref{Day}), which are valid for all $t$.  While the  four constraints  (\ref{FFcon1})--(\ref{FFcon2}) of the system (\ref{FFev1})--(\ref{FFev4}) are differential equations involving spatial gradients, the constraints (\ref{FFqham})--(\ref{FFqcon}) of the system  (\ref{FFq1})--(\ref{FFq4}) are algebraic relations that will hold for all $t$ once they hold in an initial slice $t=t_0$, and thus can be used to set up the initial conditions in terms of the coordinates $(r,x,y)$.

As a consequence, when integrating the system (\ref{FFq1})--(\ref{FFq4}) the only spacelike constraints that need to be taken care of are (\ref{FFqham})--(\ref{FFqcon}), which are algebraic relations that can be used to set up the initial conditions in terms of the coordinates $(r,x,y)$. The evolution equations (\ref{FFq1})--(\ref{FFq4}) can be treated then as a system of ordinary differential equations constrained by a 3--parameter set of initial conditions satisfying (\ref{FFqham})--(\ref{FFqcon}) at an initial time slice. 

\subsection{Alternative scalar representations.}\label{QLValt}

The evolution equations (\ref{FFq1})--(\ref{FFq4}) are given in the representation $\{\rho_q,\,\HH_q,\,\Dm,\,\Dh\}$. However, the q--scalars $A_{q}=\{\rho_{q},\,\KK_{q},\,\HH_{q},\,\Omega_{q}\}$ are interrelated by the two constraints (\ref{FFqham}) and (\ref{Omdef}), while their fluctuations $\Delta^{(B)}=\{\Dm,\,\Dk,\,\Dh,\,\DOm\}$ are interrelated by the two constraints (\ref{FFqcon}) and (\ref{DOmdef}). Hence, it is sufficient to select any representation $\{A_q,\,\Delta^{(B)}\}$ made by any two of the $A_q$ and any two of the $\Delta^{(B)}$ (with $B\ne A$ in general) to determine completely the dynamics of LTB models through evolution equations analogous (and equivalent) to (\ref{FFq1})--(\ref{FFq4}). Considering the relation between $\HH_q,\,\Omega_q$ and cosmological observable parameters, a useful alternative representation given by $\{\Omega_q,\,\HH_q,\,\DOm,\,\Dh\}$ follows by using (\ref{FFqham})--(\ref{DOmdef}) to eliminate $\rho_q$ and $\Dm$ in terms of $\Omega_q$ and $\DOm$, leading to
\bse\ba
\dot \HH_q &=& -\left(1+\frac{1}{2}\Omega_q\right)\,\HH_q^2,\label{FFq21}\\
\dot\Omega_q &=& -\Omega_q\,(1-\Omega_q)\,\HH_q,\label{FFq22}\\
\dot\Dh &=& -\left[\left(1+3\Dh\right)\Dh+\frac{1}{2}\Omega_q\left(\Dh+\DOm\right)\right]\,\HH_q, \label{FFq23}\\
\dot\DOm &=& \left[\Omega_q\DOm+\left(\Omega_q-3\DOm-1\right)\Dh\right]\,\HH_q, \label{FFq24}
 \ea\ese
As the system (\ref{FFq1})--(\ref{FFq4}), these evolution equations can also be treated as a system of autonomous ODE's subjected to the same algebraic constraints. However, equations (\ref{FFq21})--(\ref{FFq24}) are more suited to  be used to generate an dynamical systems study for Szekeres models as has been done with the LTB model with and without a cosmological constant \cite{ds1,ds2}.

\section{Initial conditions.}\label{Initcond}

Since $\{\rho_q,\,\HH_q,\,\KK_q,\,\hOm\}$ do not depend on $(x,y)$, they become $r$-dependent functions $\{\rho_{q0},\,\HH_{q0},\,\KK_{q0},\,\hOmi\}$ in an initial slice $\T[t_0]$ for an arbitrary $t_0$ (the subindex ${}_0$ will denote henceforth evaluation at $t=t_0$). On the other hand, the initial value forms for the relative fluctuations $\{\Dim,\,\Dik,\,\Dih,\,\DiOm\}$ will depend on $(r,x,y)$ through the function $\EE$ in (\ref{Edef}), whose gradient $\EE'/\EE$ enters in the definitions of these fluctuations in (\ref{Dadef}), which are valid for all $t$. Specifying $\EE$ requires prescribing the three arbitrary $r$-dependent functions $S,\,P,\,Q$. 

Because of the constraint (\ref{FFqham}), we can choose any two of the four functions $\{\rho_{q0},\,\HH_{q0},\,\KK_{qi},\,\hOmi\}$ as initial value functions. These two initial value functions, together with $S,\,P,\,Q$, are sufficient to determine the whole set of initial conditions for evolution equations like (\ref{FFq1})--(\ref{FFq4}) or  (\ref{FFq21})--(\ref{FFq24}) by means of (\ref{Dadef}), (\ref{FFqham}), (\ref{FFqcon}) (\ref{Omdef}) and (\ref{DOmdef}). It is useful to fix the $r$ coordinate also in the initial slice by the choice~\footnote{This coordinate choice is not appropriate if the slices $\T[t]$ have spherical ($\mathbb{S}^3$) or wormhole ($\mathbb{S}^2\times \mathbb{R}$) topologies. In these cases, $R_0(r)$ must have two zeroes or no zeroes. We look at these cases in \ref{wormholes}.}
\begin{equation}\fl
R_0 = r,\qquad \hbox{so that}\qquad Y_0 = \frac{r}{\EE},\quad \frac{r\,Y'_0}{Y_0}=1-\frac{r\EE'}{\EE}.\label{Ridef}
\end{equation}
As an example, if we choose the set of initial value functions $ \{\rho_{q0},\,\KK_{q0},\,S,\,P,\,Q\}$, the remaining initial value functions needed to integrate (\ref{FFq1})--(\ref{FFq4}) are
\bse\ba
 \HH_{q0}^2 =\frac{8\pi}{3}\rho_{q0}-\KK_{q0},\qquad \hOmi = \frac{8\pi\rho_{q0}}{\HH_{q0}^2},\label{ic11}\\
 \Da_0 = \frac{\delta_0^{(A)}}{1-r\EE'/\EE},\qquad A = \rho,\,\KK,\,\HH,\,\Omega,\label{ic12}\ea\ese
with the $\delta_0^{(A)}$ being the LTB initial fluctuations obtained from (\ref{Dadef}) at $t=t_0$ for LTB models ($\EE'=0,\,\,Y=R$):
\bse\ba \Diim = \frac{r}{3}\frac{\rho'_{q0}}{\rho_{q0}},\qquad \Diik = \frac{r}{3}\frac{\KK'_{q0}}{\KK_{q0}},\label{ic13}\\
\Diih = \frac{r}{3}\frac{\HH'_{q0}}{\HH_{q0}}=\frac{\hOmi}{2}\,\Diim+\frac{1-\hOmi}{2}\,\Diik,\label{ic14}\\
\DiiOm = \frac{r}{3}\frac{\hOmi'}{\hOmi}=(1-\hOmi)(\Diim-\Diik),\label{ic15}\ea\ese
where we used (\ref{Dadef}), (\ref{FFqham}), (\ref{FFqcon}), (\ref{Omdef}) and (\ref{DOmdef}).
For the alternative system (\ref{FFq21})--(\ref{FFq24}) the appropriate choice of initial value functions is furnished by the set $\{\HH_{qi},\,\hOmi,\,S,\,P,\,Q\}$. This is a practically useful alternative since these functions can be related to the Hubble an Omega parameters at $t=t_0$. In this case, the remaining functions are 
\bse\ba
\fl \frac{8\pi}{3}\rho_{q0} = \HH_{q0}^2\,\hOmi,\qquad \KK_{q0} = \HH_{q0}^2\,(\hOmi-1),\qquad \Dih = \frac{r\,\HH'_{q0}/\HH_{q0}}{3\,(1-r\EE'/\EE)},\label{ic21}\\
\fl \Dim = \frac{\DiiOm+2\Diih}{1-r\EE'/\EE},\qquad \Dik = \frac{[\hOmi/(\hOmi-1)]\,\DiiOm+2\Diih}{1-r\EE'/\EE},\label{ic22}
\ea\ese
The relation between these initial functions and the standard free parameters $M$ and $K$ follows readily from (\ref{rhoq}), (\ref{RRq}) and (\ref{Ridef}):
\begin{equation}\fl 2M = \frac{8\pi}{3}\rho_{q0}\,r^3 = \hOmi\HH_{q0}^2\,r^3,\qquad K = \KK_{q0}\,r^2=(\hOmi-1)\HH_{q0}^2\,r^2,\label{MK}\end{equation}
while the bang time $\tbb$ can be obtained as a function of the initial value functions (for example $\rho_{q0},\,\KK_{q0}$) from the solutions of (\ref{efe1}) (see section 6.2).

In order to prescribe the ``non--spherical'' part of the initial conditions it is useful to transform the $(x,y)$ coordinates of (\ref{szmetric}) by means of a stereographic projection of polar coordinates \cite{HK02,HK08}, which takes the following form:
\begin{equation}
\fl \{ x - P, \,y-Q \} = \left\{ \begin{array}{lll}
\left\{S\, {\rm cot} \left( \theta/2 \right) \cos (\phi), \,S\, {\rm cot} \left( \theta/2 \right) \sin (\phi) \right\} & \mbox{for $\epsilon = 1$},\\
\\
\left\{ S  \left( \theta/2 \right) \cos (\phi), \,S \left( \theta/2 \right) \sin (\phi) \right\} & \mbox{for $\epsilon = 0$},\\
\\
\left\{ S {\rm coth} \left( \theta/2 \right) \cos (\phi), \,S {\rm coth} \left( \theta/2 \right) \sin (\phi) \right\} & \mbox{for $\epsilon = -1$},
\end{array} \right.
\label{angcoords3}
\end{equation}
and leads to a non-diagonal metric because the transformation also depends on $r$ through $S,\,P,\,Q$ (see \cite{Bsz1,Bsz2}). 
However, this has no consequence for our purposes which is to parametrize an appropriate domain of initial conditions by means of angular coordinates $(\theta$, $\phi)$ with finite ranges and clear geometric interpretation. Applying (\ref{angcoords3}) 
to (\ref{Edef}), we can rewrite ${\cal E}'/   {\cal E}$ as \cite{HK02,HK08}
\begin{equation}
\frac{\EE'}{\EE} = \left\{ \begin{array}{lll}
-\left[S' \cos \theta + \sin \theta \left( P' \cos \phi + Q'
\sin \phi \right)\right]/S & \mbox{for $\epsilon = 1$},\\
\\
-\left[S'  \theta \left( P' \cos \phi + Q'
\sin \phi \right)\right]/S & \mbox{for $\epsilon = 0$},\\
\\
-\left[S' \cosh \theta + \sinh \theta \left( P' \cos \phi + Q'
\sin \phi \right)\right]/S & \mbox{for $\epsilon = -1$},
\end{array} \right.
\label{nuz1}
\end{equation}
so that the particular case with axial symmetry ($\epsilon = 1$ and $Q,\,P$ constants) yields
\begin{equation}
\frac{\EE'}{\EE} = -\frac{S'}{S}\,\cos \theta. \label{nuz2}
\end{equation}
which clearly has a dipolar form\footnote{Note that this is only one particular representation of the axially symmetric
case. Other forms that include P and Q functions are also possible.
This representation, however, has the simplest structure. For other
parameterizations  of the axially symmetric cases see \cite{BKHC2009}.}.

Since practically all relevant expressions associated with the Szekeres models are formally identical with the corresponding LTB expressions, save for the presence of the term $1-r\EE'/\EE$ (which is multiplied or added), then it is very useful to employ (in numerical and analytic computations) its form (\ref{nuz1}) in angular coordinates, or (\ref{nuz2}) for the case with axial symmetry, even if these quantities have been obtained in the standard $(x,y)$ coordinates. However, we will use angular coordinates only to calculate $\EE'/\EE$ in setting up initial conditions, with the time evolution obtained from systems like (\ref{FFq1})--(\ref{FFq4}) or (\ref{FFq21})--(\ref{FFq24}).

We emphasize that the initial conditions (\ref{ic11})--(\ref{ic12}) (or (\ref{ic21})--(\ref{ic22})) have been obtained by applying the algebraic constraints (\ref{FFqham})--(\ref{FFqcon}) of the system, while the constraints involving spatial gradients of the fluid flow system ((\ref{FFcon1})--(\ref{FFcon4})) correspond to mathematical identities satisfied by the fluctuations $\Da$ ((\ref{Dadef}) and (\ref{Dax})--(\ref{Day})) that are satisfied for all $t$, and thus are satisfied at $t=t_0$. As a consequence, given any set of initial value functions (which must include $S,\,P,\,Q$), the dynamics of any quasi--spherical Szekeres model can be fully determined by the  system (\ref{FFq1})--(\ref{FFq4}), which can  be integrated numerically by means of techniques used for autonomous ODE's.

\section{An initial value framework.}\label{Initval}

Together with the evolution equations described before, we can examine the dynamics of Szekeres models by analytic expressions given in terms of scaling laws with respect to a given set of initial conditions (as in \cite{RadAs,RadProfs} with LTB models). Evidently, these scaling laws are the analytic solutions of the evolution equations. 

We define the dimensionless scale factors
\ba a = \frac{Y}{Y_0} = \frac{R}{R_0} = \frac{R}{r},\label{Ldef}\\
\tilde \Gamma = \frac{Y'/Y}{Y'_0/Y_0}=\frac{\Gamma -r\EE'/\EE}{1-r\EE'/\EE},\qquad\hbox{with:}\quad \Gamma = \frac{R'/R}{R'_0/R_0}=1+\frac{r\,a'}{a},\label{Gdef}\ea
where we used the coordinate choice (\ref{Ridef}). The metric (\ref{szmetric}) takes the FLRW--like form
\begin{equation}
\fl \dd s^2 = -c^2\dd t^2 + a^2\,\left[\frac{\left(\Gamma-r\EE'/\EE\right)^2\,\dd r^2}{\epsilon-\KK_{q0} r^2}+\frac{r^2(\dd x^2+\dd y^2)}{\EE^2}\right].
\end{equation}

\subsection{Scaling laws.}\label{Slaws}

Since $M=M(r)$ and $K=K(r)$, considering (\ref{rhoq}), (\ref{RRq}), (\ref{notation}) and (\ref{MK}), and comparing the expressions for $\rho$ and $\KK$ in (\ref{locvars1}) with the expressions for $\rho$ and $\RR$ in (\ref{efe2}) and (\ref{RR1}), together with (\ref{tildeMK}) and (\ref{Dadef}), we obtain  the following scaling laws in the representation $\{\rho,\,\KK,\,\Dm,\,\Dk\}$ 
\bse\ba \rho_q=\frac{\rho_{q0}}{a^3},\quad \KK_q=\frac{\KK_{q0}}{a^2},\quad \label{slaw_mk}\\
1+\Dm = \frac{1+\Dim}{\tilde \Gamma} = \frac{1+\Diim-r\EE'/\EE}{\Gamma-r\EE'/\EE},\label{slaw_Dm}\\
\frac{2}{3}+\Dk = \frac{2/3+\Dik}{\tilde \Gamma} = \frac{\Diik+(2/3)(1-r\EE'/\EE)}{\Gamma-r\EE'/\EE},\label{slaw_Dk}
\ea\ese
where $\tilde\Gamma$ and $\Gamma$ are defined in (\ref{Gdef}) and $\Diim,\,\Diik$ are the LTB initial fluctuations given by (\ref{ic13})--(\ref{ic14}). The scaling laws for the remaining q--scalars and fluctuations follow readily from (\ref{Thetaq}), (\ref{FFqham}), (\ref{FFqcon}), (\ref{Omdef}) and (\ref{DOmdef}):
\ba \HH_q^2=\frac{\dot a^2}{a^2}=\frac{2m_{q0}}{a^3}-\frac{\KK_{q0}}{a^2},\label{slaw_H}\\
\hOm = \frac{2m_{q0}}{2m_{q0}-\KK_{q0} a},\qquad \hOm - 1 = \frac{\KK_{q0}\,a}{2m_{q0}-\KK_{q0} a},\label{slaw_Om}\ea
where $2m_{q0}\equiv (8\pi/3)\rho_{q0}$ and
\ba
\Dh = \frac{\Omega_{q0}\left(1+\Diim-\Gamma\right)+(1-\Omega_{q0})\,a\left[\Diik+\frac{2}{3}(1-\Gamma)\right]}{2\left[\Omega_{q0}+(1-\Omega_{q0})\,a\right]\,\left(\Gamma -r\EE'/\EE\right)},\label{slaw_Dh}\\
\DOm = \frac{(1-\Omega_{q0})\,a\,\left[\Diim-\Diik+\frac{1}{3}(1-\Gamma)\right]}{\left[\Omega_{q0}+(1-\Omega_{q0})\,a\right]\,\,\left(\Gamma -r\EE'/\EE\right)}\label{slaw_DOm}.\ea  
It follows from these scaling laws that for all models $\rho_q,\,\KK_q,\,\HH_q$ diverge as $a\to 0$, while $\hOm\to 1$ holds in this limit. On the other hand, all the $\Da$ diverge as $\tilde\Gamma\to 0$ (or $\Gamma\to r\EE'/\EE$), which marks a shell crossing singularity if it occurs for $a> 0$. 

\subsection{Analytic solutions.}\label{Ansols}

Implicit analytic solutions of the quadrature (\ref{quadrature}) in the representation $\{\rho,\,\KK,\,\Dm,\,\Dk\}$ depend on the sign of $\KK_{q0}$, leading to parabolic ($\KK_{q0}=0$), hyperbolic ($\KK_{q0}\leq 0$) and elliptic ($\KK_{q0}\geq 0$) models 
\ba \fl \hbox{Parabolic},\qquad t-\tbb = \frac{2 a^{3/2}}{3\sqrt{2m_{q0}}},\label{ctpar}\\
\fl \hbox{Hyperbolic},\qquad t-\tbb = \frac{Z_h(\alpha_0\, a)}{\beta_0},\label{cthyp}\\
\fl \hbox{Elliptic},\qquad t-\tbb  = \left\{ \begin{array}{l}
 Z_e(\alpha_0\, a)/\beta_0 \qquad\qquad \hbox{expanding phase}:\,\HH_q>0,\\ 
  \\ 
 \left[2\pi-Z_e(\alpha_0\, a)\right]/\beta_0 \qquad \hbox{collapsing phase}:\,\HH_q<0,\\ 
 \end{array} \right.\label{ctell}
\ea
where $\alpha_0=|\KK_{q0}|/m_{q0},\,\beta_0=|\KK_{q0}|^{3/2}/m_{q0}$ and $Z_h$ and $Z_e$ are
\bse\ba
  u  \mapsto Z_h(u)=u ^{1/2} \left( {2 + u } \right)^{1/2}  - \hbox{arccosh}(1 + u ),\label{hypZ1a}\\
 u  \mapsto  Z_e(u)= \arccos(1 - u )-u ^{1/2} \left( {2 - u } \right)^{1/2}.\label{ellZ1a} 
\ea\ese
Since $a=1$ at $t=t_0$ and $a=0$ at $t=\tbb$ and $t=\tcoll$ (for elliptic models), we obtain the bang and collapse time functions
\bse\ba \fl \tbb = t_0 -\frac{2}{3\sqrt{2m_{q0}}}\qquad \hbox{Parabolic},\label{tbbp}\\
\fl \tbb = t_0-\frac{Z_h(\alpha_0)}{\beta_0},\qquad \hbox{Hyperbolic},\label{tbbh} \\
\fl \tbb = t_0-\frac{Z_e(\alpha_0)}{\beta_0},\quad \tcoll = \tbb+\frac{2\pi}{\beta_0}=t_0+\frac{2\pi-Z_e(\alpha_0)}{\beta_0},\qquad\hbox{Elliptic},\label{tbbe}\ea\ese
Notice that it is possible to use $\tbb$ as an initial value function. This requires providing a specific functional form for $\tbb$ and (say) $\rho_{q0}$, then we can find $\KK_{q0}$ by solving (\ref{tbbh}) or (\ref{tbbe}) numerically. Also, the case with a simultaneous big bang transforms (\ref{tbbh}) and (\ref{tbbe}) into constraints so that given $m_{q0}=(4\pi/3)\rho_{q0}$ we find $\KK_{q0}$ by solving them with $\tbb =\tbb^{(0)}=$ constant.

Note that the above relations do not depend on $(x,y)$ variables and are
the same for the quasi-spherical, quasi-hyperbolic, and quasi-plane cases.
The dependence on $(x,y)$ variables enters only via initial conditions $\Dim$ and/or $\Dik.$

By implicit derivation of the solutions (\ref{ctpar}), (\ref{cthyp}) and (\ref{ctell}) we obtain the following form for $\tilde\Gamma$ given by (\ref{Gdef}): 
\bse\ba \fl \hbox{Parabolic}\qquad \tilde\Gamma = 1+\Dim-\frac{\sqrt{2m_{q0}}\,r\,\tbb'}{a^{3/2} (1 - r\EE'/\EE)},\label{Gp}\\
\fl \hbox{Hyperbolic and elliptic}\nonumber\\
\fl \tilde\Gamma =  1+3(\Dim-\Dik)-3\left(\Dim-\frac{3}{2}\Dik\right)\HH_q (t -\tbb)-\frac{r\HH_q \tbb'}{1-r\EE'/\EE},\label{Ghe}\ea\ese 
where in the hyperbolic and elliptic case $\tbb$ is given by (\ref{tbbh}) and (\ref{tbbe}) and 
\begin{equation}
\fl \frac{r\,\tbb'}{3(1-r\EE'/\EE)} = \frac{\Dim-\Dik}{\HH_{q0}}-\left(\Dim-\frac{3}{2}\Dik\right)\,(t_0-\tbb),\label{tbbr}
\end{equation}
follows from deriving (\ref{tbbh}) and (\ref{tbbe}) and using (\ref{Dai}). All local scalars $\{\rho,\,\KK,\,\HH,\,\Sigma,\,\W\}$ can now be given as functions of $a$ and the initial value functions by applying the scaling laws (\ref{slaw_mk})--(\ref{slaw_Dh}) to the relations (\ref{locvars1})--((\ref{locvars2}), with $\tilde\Gamma$ given by (\ref{Gp}) or (\ref{Ghe}).

\subsection{Scaling laws and analytic solutions in other representations.}\label{Repr}

 \subsubsection{The representation $\{\Omega_q,\,\HH_q,\,\Dm,\,\Dk\}$}.
The scaling laws and analytic solutions follow from the  expressions derived before by replacing $\rho_{q0}$ and $\KK_{q0}$ with $\Omega_{q0}$ and $\HH_{q0}$ by means of (\ref{ic21})--(\ref{ic22}), so that parabolic, hyperbolic and elliptic models now correspond respectively to $\hOmi-1=0,\,\hOmi-1<0$ and $\hOm-1>0$. The scaling laws (\ref{slaw_mk}), (\ref{slaw_H}) and (\ref{slaw_Om}) take FLRW forms
\bse\ba \frac{8\pi}{3}\rho_q =\frac{\Omega_{q0}\HH_{q0}^2}{a^3},\qquad \KK_q = \frac{(\Omega_{q0}-1)\HH_{q0}^2}{a^2},\\
\HH_q^2 = \HH_{q0}^2\left[\frac{\hOmi}{a^3}+\frac{1-\hOmi}{a^2}\right],\label{OmH2}\\
\hOm =\frac{\hOmi}{\hOmi+(1-\hOmi)\,a},\qquad \hOm-1=\frac{(\hOmi-1)\,a}{\hOmi+(1-\hOmi)\,a},\label{OmH22}\ea\ese
while the fluctuations $\Da$ take the same form as in (\ref{slaw_Dm}), (\ref{slaw_Dk}), (\ref{slaw_Dh}) and (\ref{slaw_DOm}). The analytic solutions (\ref{ctpar})--(\ref{ctell}) and the forms of $\tbb$ in (\ref{tbbp})--(\ref{tbbe}), as well as previous expressions for $\tilde\Gamma$ and $\tbb'$ have the same forms with $\alpha_0$ and $\beta_0$ given by:
\begin{equation}\alpha_0 = \frac{2\hOmi}{|\hOmi-1|},\qquad \beta_0 = \frac{2\hOmi\,\HH_{q0}}{|\hOmi-1|^{3/2}}. \end{equation}
\subsubsection{The representation $\{\Omega_q,\,\HH_q,\,\DOm,\,\Dh\}$}.
If we keep $a$ as scale factor then the scaling laws for the q--scalars and their fluctuations are the same as in the representation $\{\Omega_q,\,\HH_q,\,\Dm,\,\Dk\}$ above, with $\Dim,\,\Dik$ expressed in terms of $\Dih,\,\DiOm$ by (\ref{ic22}).
 
Another possibility follows by using $\Omega_q$ as scale factor by eliminating $a$ and $\HH_q$ in terms of $\HH_{q0},\,\hOmi,\,\hOm$ by means of
\bse\ba a =\frac{\hOmi\,(1-\hOm)}{\hOm\,(1-\hOmi)},\label{OmH1}\\
\HH_q = \HH_{q0}\,\frac{\hOm}{\hOmi}\,\left[\frac{1-\hOmi}{1-\hOm}\right]^{3/2}. \label{OmH3} \ea\ese
The analytic solutions of the quadrature (\ref{quadrature}) become trivial for the parabolic case ($\Omega_q=1$), while for the hyperbolic and elliptic case they take the form
\ba \fl \hbox{hyperbolic}\quad 0<\Omega_{q0}<1:\qquad 
 t-t_0 =\frac{W-W_{0}}{\HH_{q0}},\label{hypsol2}\\
 \fl \hbox{elliptic}\quad \Omega_{q0}>1:\nonumber\\
\fl t - t_0  = \left\{ \begin{array}{l}
 [W-W_{0}]/\HH_{q0} \quad
\hbox{(expanding phase)},\\  
 \left[\pi\Omega_{q0}(\Omega_{q0}-1)^{-3/2} - W-W_{0}\right]/\HH_{q0}\quad 
  \hbox{(collapsing phase)},\\
 \end{array} \right.\label{ellsol2},\ea
where the functions $W$ and $W_0=W|_{\Omega_q=\Omega_{q0}}$ take the form
\ba \fl  W=\frac{\varepsilon_0\Omega_{q0}}{2\,|1-\Omega_{q0}|^{3/2}}\,\left[\frac{2\sqrt{\varepsilon_0(1-\Omega_q})}{\Omega_q}-\ACal\left(\frac{2}{\Omega_q}-1\right)\right], \label{W}\\
 \fl W_0=\frac{\varepsilon_0}{|1-\Omega_{q0}|}\,\left[1-\frac{\Omega_{q0}}{2\,|1-\Omega_{q0}|^{1/2}}\,\ACal\left(\frac{2}{\Omega_{q0}}-1\right)\right],\label{W0} \ea
where $\varepsilon_0 =1,\, \ACal=$ arccosh correspond to the hyperbolic case ($\Omega_{q0}<1$) and $\varepsilon_0 =-1,\, \ACal=$ arccos to the elliptic case ($\Omega_{q0}>1$). The bang time is given by
\begin{equation}\tbb = t_0-\frac{W_0(\Omega_{q0})}{\HH_{q0}},\label{tbb2}\end{equation}
while $\Gamma$ and the gradient of the bang time follow from (\ref{Ghe}) and (\ref{tbbr}) with $\HH_q$ given by (\ref{OmH3}) and $\Dim,\,\Dik$ eliminated on terms of $\DiOm,\,\Dih$ by means of (\ref{ic22}). 

Using the functions $\HH_q,\,\hOm$ and their initial values as variables of the scaling laws can be very useful for future work in applying Szekeres models to fit observations, as $\HH_q,\,\hOm$ provide an appealing generalization of FLRW cosmological observational parameters (they reduce to these parameters in the FLRW limit). In fact, LTB void models used to fit observations are often parametrized in terms of  $\HH_q,\,\hOm$ exactly defined as in (\ref{OmH1}) and (\ref{OmH2}), which are introduced as ansatzes \cite{moffat,kolb,alnes,endqvist,kbsn1,GBH,kbsn2,clarkson}. Since $\HH_q$ is a QL variable and $\hOm$ is the ratio of QL variables, and the latter are Szekeres variables independent of the ``non--radial'' coordinates ${x,y}$, then it is expected that they exactly correspond to (and satisfy the same relations) as their analogous LTB variables.

\section{Regularity Conditions.}\label{Regconds}

As with LTB models, Szekeres models admit two types of curvature singularities: a ``central'' singularity associated with $a=0$ and $t=\tbb$ or (in elliptic models) $t=\tcoll$, and a shell crossing singularity $\tilde\Gamma(t,r,x,y)=0$. Notice that the ``non--radial'' variables $(x,y)$ play no role in determining the locus of $a=0$, but they are involved in the shell crossing singularity (and in the conditions to avoid it). Also, it is important to remark that  $a=0\,\,\Rightarrow\,\, Y=0$ and $\tilde \Gamma=0\,\,\Rightarrow\,\, Y'=0$, but the converses of these implications are not true.

The condition to avoid a shell crossing singularity is given by $\tilde\Gamma>0$ and by using (\ref{Gp}) and (\ref{Ghe}) it can be expressed as conditions on the initial value functions (the generalization of Hellaby--Lake conditions in LTB models \cite{ltbstuff1,ltbstuff2}). 
Particular conditions
depend whether we have the parabolic, hyperbolic and elliptic cases:
\begin{description}
\item[Parabolic models.] It is evident from (\ref{Gp}) that the necessary and sufficient condition for $\tilde\Gamma>0$ is given by:
\begin{equation}
-1\leq \Dim \leq 0,\label{HLp}
\end{equation}
where we used the relation between $\Dim$ and $\Diim$ in (\ref{ic13}) together with $\Diim = \HH_{q0}\,r\tbb'$. Notice that the condition involving $\Dim$ is equivalent to $\tilde M'\geq 0$ and $\rho_0=\rho_{q0}[1+\Dim]\geq 0$, where $\tilde M$ is defined by (\ref{tildeMK}) and $\rho_0$ is the initial local density\footnote{ $\tilde M'\geq 0$ implies $M'- M \EE'/EE \geq 0$,
this means that unlike in the LTB model we cannot have $M\approx$ constant.}.
 Conditions (\ref{HLp}) reduce to their equivalent forms for parabolic LTB models ($\EE'=0$) \cite{suss2010a,RadAs,RadProfs,ltbstuff1,ltbstuff2}.
\item[Hyperbolic models.] We look at the form of $\tilde \Gamma$ in (\ref{Gdef}) in the following asymptotic limits along comoving worldlines:
\begin{itemize}
\item  $a\to 0,\quad \HH_q\to\infty,\quad \HH_q (t-\tbb) \approx \frac{2}{3}+O(a)$
\begin{equation}\tilde\Gamma \approx 1+\Dim -r\HH_q\, \tbb'/(1-r\EE'/\EE),\label{Glimtbb}\end{equation}
\item $a\to\infty,\quad \HH_q\to 0,\quad \HH_q (t-\tbb) \approx 1+O\left(\frac{\ln a}{a}\right)$
\begin{equation} \tilde\Gamma \approx 1+\frac{3}{2}\Dik.\label{Gliminf} \end{equation}
\end{itemize}
Hence, necessary and sufficient conditions for $\tilde\Gamma>0$ are given by:
\begin{equation}
\Dim \geq -1,\quad \Dik \geq -\frac{2}{3},\quad \frac{\tbb'}{1-r\EE'/\EE}\leq 0,\label{HLh}\end{equation}
As in the parabolic case, the condition given in terms of $\Dim$ implies that $\tilde M'$ and $\rho_0$ are non--negative, while the condition given in terms of $\Dik$ implies that $\tilde K'\geq 0$ (notice that $\tilde K\leq 0$ and $\KK_{q0}\leq 0$ hold for hyperbolic models). Conditions (\ref{HLh}) reduce to their forms for LTB models if $\EE'=0$ so that $\Dim=\Diim$ and $\Dik=\Diik$ hold \cite{suss2010a,ltbstuff1,ltbstuff2,RadAs,RadProfs}.
\item[Elliptic models.] At the surface of maximal expansion ($\HH_q=0$), we have from (\ref{Gdef}) 
\begin{equation}
\tilde\Gamma= 1+3(\Dim-\Dik).\label{Gmax}
\end{equation}
In the limit $a\to 0$ with $t\to\tbb$ we obtain the same expression (\ref{Glimtbb}) as in the hyperbolic case in the same limit, but in the limit $a\to 0$ with $t\to\tcoll$ we have $\HH_q\to-\infty$, hence we obtain
\begin{equation}
\tilde\Gamma \approx 1+3(\Dim-\Dik)+\frac{r\,|\HH_q|\,\tcoll'}{1-r\EE'/\EE},\label{Glimtcoll}
\end{equation}
where
\begin{equation}
\fl \frac{r\,\tcoll'}{1-r\EE'/\EE} = 3\left(\Dim-\frac{3}{2}\Dik\right)\,(\tcoll-\tbb)+\frac{r\,\tbb'}{1-r\EE'/\EE},\label{tcollr}
\end{equation}
follows from (\ref{tbbe}). Hence, considering (\ref{Glimtbb}) and (\ref{Gmax})--(\ref{tcollr}), we have the following conditions for $\tilde\Gamma>0$ in terms of Szekeres initial fluctuations  
\bse\ba \fl 1+3(\Dim-\Dik)>0,\quad \Dim-\frac{3}{2}\Dik>0,\qquad \hbox{necessary}, 
\label{SCcond1} \\
\fl \Dim \geq -1,\quad \frac{\tbb'}{1-r\EE'/\EE}\leq 0,\quad \frac{\tcoll'}{1-r\EE'/\EE}\geq 0,\;\;\hbox{necessary and sufficient}, \label{SCcond2} \ea\ese
As in the parabolic and hyperbolic cases, these conditions reduce to the Hellaby--Lake conditions in the LTB limit $\EE'=0,\,\Da_0=\Daa_0$ \cite{suss2010a,RadAs,RadProfs,ltbstuff1,ltbstuff2}. 

\end{description} 

\section{Comparison with LTB models.}\label{CompLTB}

Since LTB models are well known inhomogeneous cosmological models that have been frequently utilized, it is useful and desirable to compare them with the quasi--spherical Szekeres models, which provide a straightforward non--spherical generalization. 

The evolution equations (\ref{FFq1})--(\ref{FFq4}) are identical to those of an LTB model \cite{suss2010a}, save for the fact that $\Dm$ and $\Dh$ depend on $(x,y)$, though this dependence only needs to be prescribed as part of the initial conditions through the term $1-r\EE'/\EE$, with $\EE$ given by (\ref{Edef}) or $\EE'/\EE$ by (\ref{nuz1}) in spherical coordinates. Likewise, the forms of the QL scalars $\rho_q,\,\KK_q, \, \HH_q$ and $\hOm$ given by the scaling laws (\ref{slaw_mk})--(\ref{slaw_Om}) are identical to those of an LTB model described by the q--scalars and their fluctuations \cite{suss2010a,RadAs,RadProfs} (notice that the determination of the scale factor $a$ in (\ref{ctpar})--(\ref{ctell}) does not require that $\EE$ is specified). 

Since the q--scalars are common to both LTB and Szekeres models, the difference between these models only enters in the fluctuations through the term $1-r\EE'/\EE$, as can be appreciated from the relation between initial fluctuations $\Da_0$ and $\Daai$ of Szekeres and LTB models given in (\ref{ic13})--(\ref{ic14}): 
\begin{equation}
\Da_0 =  \frac{\delta_0^{(A)}}{1-r\EE'/\EE},\qquad \Daai = \frac{rA'_{q0}}{3A_{q0}},\qquad A = \rho,\,\KK,\,\HH,\,\Omega,\label{Dai}
\end{equation}
Considering the fact that any LTB model can be characterized as a unique solution of the system (\ref{FFq1})--(\ref{FFq4}) for initial value functions $\{\rho_{q0},\,\KK_{q0},\,\Diim,\,\Diik\}$, then each LTB model can be associated with a Szekeres model by the following transformation in the space of initial conditions:
\begin{equation}\Daa_0 \mapsto \Da_0,\qquad \hbox{so that}\qquad \Daa \mapsto \Da,\label{LTBtoSz}\end{equation}
with $\Da$ for $A=m,\,k,\,\HH$ given by (\ref{slaw_Dm}), (\ref{slaw_Dk}), (\ref{slaw_Dh}) and (\ref{DOmdef}). 
The transformation (\ref{LTBtoSz}) simply requires modifying the initial fluctuations of an arbitrary LTB model by choosing (besides $\rho_{q0},\,\KK_{q0}$) the three extra free functions $\{S,\,P,\,Q\}$ to construct the term $\EE'/\EE$ in (\ref{nuz1}). Szekeres models obtained in this manner form a 3--parameter class of models associated with a unique LTB model that follows from the solution of the same evolution equations ({\it i.e.} (\ref{FFq1})--(\ref{FFq4})) but with the modified initial conditions $\{\rho_{q0},\,\KK_{q0},\,\Dim,\,\Dik\}$. Conversely, any quasi--spherical Szekeres model can be mapped to a unique LTB model with the same QL scalars $\rho_q,\,\KK_q,\,\HH_q$ and $\hOm$ (given by (\ref{slaw_mk})--(\ref{slaw_Om})) and fluctuations transformed by (\ref{LTBtoSz}).

In particular, the relation between a given LTB model and the 3--parameter class of associated Szekeres models can be understood in terms of a perturbative approach if we choose the free functions $\{S,\,P,\,Q\}$ so that $r\EE'/\EE\ll1 $, which implies:
\begin{equation}
\Da_0 \approx \left(1+\frac{r\EE'}{\EE}\right)\,\delta_0^{(A)},
\label{dipper}
\end{equation}
so that the initial fluctuations $\Da_0$ take the form of perturbations of the LTB fluctuations. Under these conditions, we can examine Szekeres models that are {\it almost} LTB, with perturbative deviations from spherically symmetric.  It is important to remark that quasi--plane and quasi--hyperbolic Szekeres models relate to dust solutions with plane and pseudo--spherical symmetry in the same manner as quasi--spherical models relate to spherically symmetric LTB solutions (which we discussed in this section).

\section{Numerical example: Growth of the dipole distribution.\label{dipole}}

\begin{figure}
\begin{center}
\includegraphics[scale=0.55]{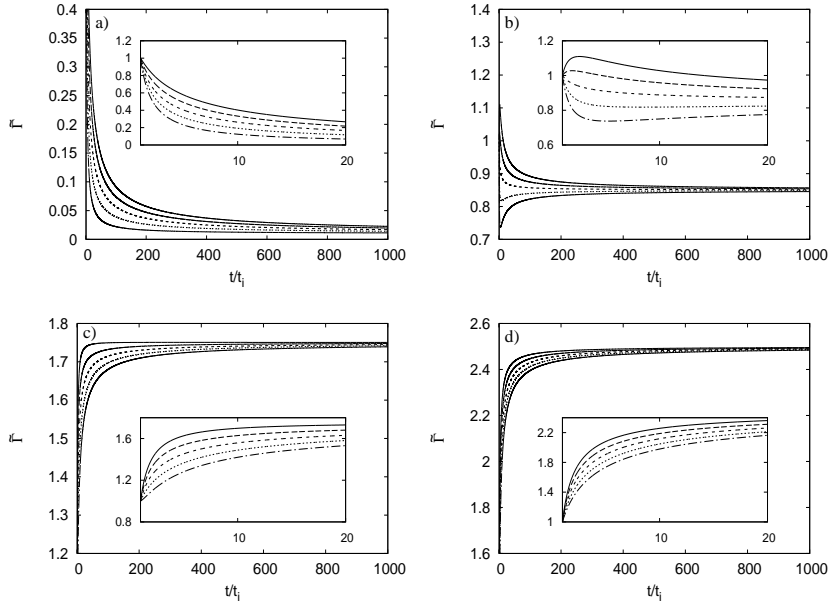}
\caption{Evolution of $\tilde{\Gamma}$ for a hyperbolic model
with negligible dipole ($r \EE '/\EE \approx 0$), and 
with $\hOm = 0.3$ and $\HH_q = 70$ km s$^{-1}$ Mpc$^{-1}$ (see (\ref{Omdef})) 
 {\em Upper left} (a) $\Diik = -0.66$, {\em upper right} (b) $\Diik = -0.1$,
{\em lower left} (c) $\Diik = 0.5$, {\em lower right} (d) $\Diik = 1$.
In each panel the curves from the top to bottom:
$\Diim = 1,0.5,0,-0.5$, and $-0.99$.}
\label{fig1}
\end{center}
\end{figure}

\begin{figure}
\begin{center}
\includegraphics[scale=0.55]{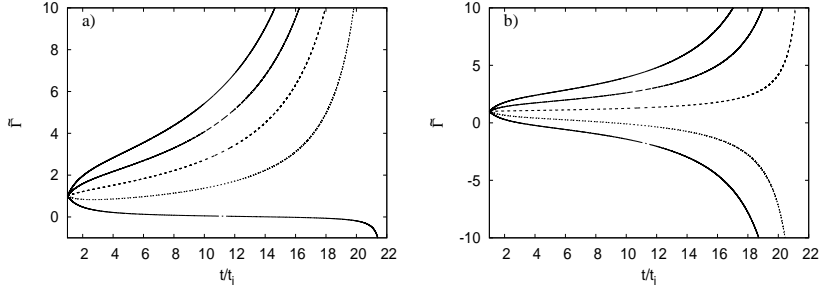}
\caption{Evolution of $\tilde{\Gamma}$ (from $t_0$
till $t_{\rm coll}$)
for a elliptic model
with negligible dipole ($r \EE '/\EE \approx 0$), and 
with $\hOm = 0.3$ and $\HH_q = 70$ km s$^{-1}$ Mpc$^{-1}$ (see (\ref{Omdef})) 
 {\em left} (a) $\Diik = -0.66$, {\em right} (b) $\Diik = -0.1$.
In each panel the curves from the top to bottom:
$\Diim = 1,0.5,0,-0.5$, and $-0.99$.}
\label{fig2}
\end{center}
\end{figure}

In order to illustrate how the theoretical framework that we have presented works in practice, 
we examine the dipole evolution that marks the deviation of a quasi--spherical Szekeres model from spherical symmetry.
In particular, we address the question of the stability of spherical symmetry with respect to dipole perturbations. Let us consider a hyperbolic ($K<0$) model with initially small deviation
from spherical symmetry, i.e. with $r \EE'/\EE \approx 0$ (see (\ref{dipper})). The evolution of this model for a 
comoving surface of fixed $r$ is calculated as follows: 

\setcounter{footnote}{0}

\begin{enumerate}
\item First we choose an FLRW background that can be identified in the asymptotic limit $r\to\infty$ of $\hOm$ and $\HH_q$ in the slice $t=t_0$. Using (\ref{Omdef}), we choose as background quantities $\Omega_0\equiv \Omega_q{}_\infty = 0.3$ and $H_0\equiv \HH_q{}_\infty = 70$ km s$^{-1}$ Mpc$^{-1}$.
\footnote{Since the examples we are considering are meant for illustrative purposes, these values simply allow us to specify a model and do not correspond to an actual model of some kind of a realistic structure.}
\item We choose the initial value functions $\rho_{q0}$ and $\KK_{q0}$ at a fixed $r$ with $t=t_0$ corresponding 
to the present cosmic time (computed from (\ref{cthyp}) considering present day values for $H_0$ and $\Omega_0$ given by the background values mentioned in point (i) above). The initial monopole perturbations $\Diim$ and $\Diik$ are computed from  
these initial condition from (\ref{Dai}).
\item We consider a small and negligible dipole perturbation. One of such choices is:  $P' = 0 = Q'$, $S = r^\alpha$ with $\alpha = 0.001$.                      
\item Knowing $\Diim$, $\Diik$ and $\EE'/\EE$  we find $\Dim$ and $\Dik$ from from (\ref{Dai}).
\item As seen from (\ref{slaw_Dm}), under these conditions the density perturbation
evolves only through the time evolution of $\tilde{\Gamma}$, which is given by 
and (\ref{Gp}) or (\ref{Ghe}). Notice that the form of $\Dm$ is not only determined by $\tilde\Gamma$, but also by the presence of $r\EE'/\EE$ in $\Dim$, 
which, however, does not depend on time.
\end{enumerate}

Since initially $\Dim \approx \Diim$ (i.e. $r\EE'/\EE \approx 0$)
thus the only possibility for a large departure from the spherical symmetry (large dipole-like fluctuations)
is when $\Gamma \approx r\EE'/\EE$.

The evolution of $\tilde{\Gamma}$ is presented in Fig. \ref{fig1}.
As seen in the cases that were considered, $\tilde{\Gamma}$ tends to an asymptotic value $\tilde{\Gamma}_{\infty}$. 
Thus, we recover the  know fact from the evolution of 
ever--expanding hyperbolic LTB models (i.e. $K<0$), that density perturbations freeze in the time 
asymptotic range \cite{PK06}, but 
only if $\tilde{\Gamma}_{\infty} \approx 0$ we can have an asymptotically large dipole variation. If $\tilde{\Gamma}_{\infty} \gg r\EE'/\EE$
then the deviation from spherical symmetry is negligible.
As can be seen from Fig. \ref{fig1}, only when $\Diik <0$ we have 
$\tilde{\Gamma}_{\infty} < 1$. Thus, if the initial deviation
from spherical symmetry is small then the spherical shape is stable as long as we exclude models that are close to shell crossing singularities (i.e. $\Diik \approx - 2/3$). We can also obtain this result by looking at the full analytic expression for the density perturbation $\Dm$ from the scaling law (\ref{slaw_Dm}) and the form of $\tilde\Gamma$ in (\ref{Ghe}) (and eliminating $\Dim,\,\Dik$ in terms of $\Diim,\,\Diik$ with (\ref{Dai})):
\begin{equation} \fl \Dm = \frac{3\left(\Diim-\frac{3}{2}\Diik\right)\left[\HH_q(t-\tbb)-\frac{2}{3}\right]+r\HH_q\tbb'}{1-r\EE'/\EE+3(\Diim-\Diik)-3\left(\Diim-\frac{3}{2}\Diik\right)\HH_q(t-\tbb)-r\HH_q\tbb'}.\label{DM}\end{equation}
In the limit $t\to\infty$ (which is equivalent to $a\to\infty$) we obtain $\HH_q\to 0$ from (\ref{slaw_H}) and $\HH_q(t-\tbb)\to 1$ from (\ref{cthyp}) and (\ref{Gliminf}). Hence, we obtain in this limit 
\begin{equation}\Dm \approx \frac{\Diim-\frac{3}{2}\Diik}{1-r\EE'/\EE+\frac{3}{2}\Diik}+
O\left(\frac{\ln a}{a}\right),\end{equation}
which shows the value at which density perturbations freeze asymptotically, and also the fact that the asymptotic effects of the deviation from spherical symmetry (contained in the term $r\EE'/\EE$) remain small if chosen to be small at $t=t_0$.

We now consider another similar numerical example. This time, however,
we take an elliptic background, thus we use the analytic solution (\ref{ctell}) and consider a value $\Omega_{q0} = 1.5>1$ for a fixed $r$.
The results are presented in Fig. \ref{fig2}.
The evolution is calculated from the initial instant $t_0$ (present cosmic time) 
all the way to the big crunch $t=t_{\rm coll}$.
As can be seen in both pannels of the figure, there are examples in which 
a shell crossing singularity occurs ($\tilde{\Gamma}=0$), and it is easy to verify that
for these values of $\Diim$ and $\Diik$ the regularity conditions (\ref{SCcond1})-(\ref{SCcond2}) are not satisfied.
If we demand that the  models are free of shell crossings, then $\tilde{\Gamma} \to 0$ must not occur, hence 
$\tilde{\Gamma} $ must increase monotonically as dust layers expand for early times and collapse for late times (there is not symmetry with respect to $t=t_{\tiny{\textrm{bounce}}}$ where $\HH_q=0$).
Thus, the density perturbation $\Dm$ monotonically decreases, and as 
$\tilde{\Gamma} \to \infty$, we have $\Dm \to -1$. Hence, a generic perturbation within the elliptic background
will either decrease ($\Dm \to -1$) or increase and eventually lead to a shell crossing.

The effects of the deviation from spherical symmetry also remain small (if originally small at $t=t_0$). This can be seen looking at the analytic form of $\Dm$ in (\ref{DM}) in the limit $t\to\tcoll$. Since $\HH_q\to-\infty$ as $t\to\tcoll$, then an asymptotic expansion of $\Dm$ around $-|\HH_q|$ yields:
\begin{equation} \Dm \approx -1 +\frac{1+\Diim-r\EE'/\EE}{|\HH_q|\,r\tcoll'}+O(|\HH_q|^{-2}),\label{DMcoll}\end{equation}
where we used (\ref{tcollr}). Evidently, the effects of the deviation from spherical symmetry (the term $r\EE'/\EE$) enter as a correction of order $1/|\HH_q|$ and thus remain small if chosen initially to be small. The limit $\Dm\to -1$ as $t\to\tcoll$ may seem strange, as one would associate a diverging density contrast near the collapsing singularity. This limiting value follows from (\ref{slaw_Dm}) and from the fact that $\Gamma$ and $\tilde\Gamma$ diverge as $\HH_q\to -\infty$ (which occurs as $t\to\tcoll$). Notice that the $\Da$ are NOT ``contrast'' perturbations, but have a more complicated interpretation related to the time evolution of the gradients $A'$ and $A'_q$ through their definition in (\ref{Dadef}). Therefore, there is no reason for $\Dm$ to diverge at $\tcoll$. The limit $\Dm\to -1$ could imply Schwarzschild vacuum conditions if $\rho\to 0$ with $\rho_q>0$ \cite{RadAs,sussBR2}, but in the collapsing regime it reflects the fact that the ratio $\rho/\rho_q$ vanishes as $t\to\tcoll$, but with both densities diverging in this limit.

The following conclusion arises from the discussion above:
if we consider only initial perturbations 
and demand absence of shell crossings together with 
an almost spherical shape with negligible dipole--like departure from spherical
symmetry, then the spherical shape is conserved.
Bearing in mind this conclusion, the following question arises: If 
spherical symmetry is a stable property, then
why do non-symmetrical structures are present in the Universe?
The reason is that even if cosmic structures (voids, clusters, superclusters) evolved
from small fluctuations that were present at the last scattering time,
these small fluctuations did not have to be almost spherical, nor the conditions
of avoidance of shell crossings had to be satisfied initially. In fact, the 
only case in which an initially large deviation from spherical 
symmetry allows us to permanently (not temporally - see insets in Fig. \ref{fig1})
dissolve the dipole occurs only when the decaying modes are present.
One example of this is a parabolic model with zero curvature
perturbations, i.e. $\KK_{q0} = 0 = \Diik$.
Then as seen from (\ref{Gp}) $\tilde{\Gamma}_{\infty} = 1+\Dim$
and so $\Dm \to 0$.

\section{Final discussion and further work.}\label{Final}

We have introduced for Szekeres models of class I a set of new coordinate independent representations of scalar variables consisting of the q--scalars and their fluctuations. We have shown throughout the article that these variables completely determine the dynamics of the models, either in terms of intuitively appealing analytically expressions (section \ref{Initval}) or through fluid flow evolution equations (section \ref{QLVeveqs}) that can be handled as ordinary differential equations (ODE's) subjected to algebraic constraints. We have also shown that by applying these constraints we can construct various equivalent representations of the new variables, some of which can be related to observational parameters such as the Hubble factor $H$ and $\Omega$ (sections \ref{QLValt} and \ref{Repr}). Various related theoretical issues have been discussed in detail: 

\renewcommand{\labelitemi}{$-$}

\begin{itemize}
\item the relation between the q--scalars and averages of covariant scalars (section \ref{QLVprops}), 

\item  initial conditions for the fluid flow evolution equations (section \ref{Initcond}), 

\item  regularity conditions to avoid shell crossings (section \ref{Regconds}), 

\item  as well as a comparison with LTB models (section \ref{CompLTB}),

\item  we have also used the new variables (section \ref{dipole}) to examine the preservation of nearly spherical initial conditions in late time regimes, and showed
 that spherical symmetry is stable against small dipole-like perturbations.
\end{itemize}

\renewcommand{\labelitemi}{$\bullet$}

We list below the main advantage of these variables over the traditional ones, as well as potential applications to be undertaken in future articles: 
\begin{description}
\item[Numerical work.]  The fluid flow evolutionary equations that were obtained in section \ref{QLVeveqs} (in any given representation) form a system of four PDE's that can be handled effectively as ordinary differential equations (ODE's). One possible representation is that given by (\ref{FFq1})--(\ref{FFq4}), consisting of the q--scalars associated with the density and expansion scalars and their fluctuations with respect to the local scalars. In another representation (in (\ref{FFq21})--(\ref{FFq24})) the four variables are the q--scalars associated with the Hubble and Omega factors and their fluctuations. In either representation, the system depends on five free parameters that convey the effects of spherical and non--spherical inhomogeneity and are specified as initial conditions. The spacelike constraints for these system are not PDE's, but algebraic equations. This represents an important advantage over the fluid flow evolution equations in terms of the local covariant scalars, which need to be handled as PDE's because the constraints are PDE's that couple in a non--trivial way to the time derivatives (as in the fluid flow systems in \cite{zibin,dunsbyetal}).  These evolution equations provide a nice and elegant (and simplified) approach to numerical work with Szekeres models, with an enormous potential for applications either in theoretical studies or in fitting observations. 

\item[Theoretical work.] By employing an initial value formalism based on the q--scalars and their fluctuations, we can

\begin{itemize}
\item  study the propagation of any given set of initial conditions, either analytically or numerically. We can compare this propagation with that of initial conditions in spherical LTB models. In particular, we can examine the stability of initial spherical shapes (LTB model)  against small (dipole like) perturbations -- see section \ref{dipole}. 
\item extend previous theoretical results obtained by means of these variables in LTB solutions to the Szekeres models. In particular, we aim in future articles to generalize previous work dealing with important features of the LTB models:  averaging inhomogeneities \cite{sussIU,LTBave1,LTBave2,sussBR1,sussBR2}, radial asymptotics \cite{RadAs}, evolution of radial profiles \cite{RadProfs},  and dynamical systems analysis \cite{ds1,ds2}.
\item extend the work done on the connection with non-linear perturbations on a FLRW background \cite{suss2009} that compares the fluctuations between local and QL variables with exact perturbations in a FLRW background, for example identifying and studying the evolution of exact quantities that generalize the growing and decaying modes of the theory of linear perturbation of dust sources.  
\end{itemize}
\end{description}

We consider that the new variables that we have introduced not only provide a deeper theoretical understanding of the Szekeres models, but is (at the same time) more intuitive than the study of the models in the  traditional variables. This formalism has an enormous potential for exploring the effects of non--spherical inhomogeneity and non--linear perturbations, it may also allow for a more efficient utilization of Szekeres solutions in for the study of cosmic inhomogeneities (including void models) to test cosmological observations. We are currently undertaking further work on these lines that we expect to submit in the near future.

\begin{appendix}

\section{Averaging}\label{averaging}

As mentioned in Section \ref{QLV}, the integral definition of the q--scalars is equivalent to a proper volume averaged integral with weight factor $\FF$. The standard  proper volume averaging (weight factor $\FF=1$) for the quasi--spherical Szekeres models was studied in \cite{Bole2009-vav}.
We show in this appendix that, save for some qualitative differences, the results and the approach of \cite{Bole2009-vav} remain valid for the quasi--hyperbolic and quasi--planar models, leading to the expressions (\ref{rhoq}), (\ref{Thetaq}) and (\ref{RRq}).  
                                                                            
First, we remark that the area of a 2--surface of constant $t$ and $r$ 
in the quasi--hyperbolic and quasi--planar models may be infinite.   
However, this is not problematic, as the surface of the domain $S_D$
cancels out. Second, a location $r=0$ that can be identified as an ``origin'' 
only exists for quasi--spherical models  --
in the quasi--hyperbolic model $r$ cannot be equal to zero,
and in the quasiplane $r$ can only asymptotically approach 
the origin, $r \rightarrow 0$ \cite{HK08}.
However, one can always consider a domain $\DD$ is centered around $r=0$
even if this point does not belong to the manifold.

For the purpose of averaging it is more  convenient to adopt a pair of complex conjugate coordinates
\begin{equation}
\zeta = x + i y, \quad \bar{\zeta} = x - i y,
\end{equation}
so that the metric (\ref{szmetric}) becomes

\begin{equation}\dd s^2 = -\dd t^2 +
 \frac{\EE^2\,Y'{}^2}{\epsilon-K}\dd r^2 + Y^2{\rm d} \zeta {\rm d} \bar{\zeta}.\label{szmetric2}\end{equation}
The 3--volume associated with the quasi--local average of the domain $\mathcal{D}$ centered at the origin is
(the weight factor $\FF = \sqrt{\epsilon-K}$, see also (\ref{qdef}))

\begin{eqnarray}
&& V_{q} =  \int\limits\limits_{r_c}^{r_{\mathcal{D}}} {\rm d} {r} \int \int {\rm d} \zeta {\rm d} \bar{\zeta} ~\EE Y' Y^2  = \int\limits_{r_c}^{r_{\mathcal{D}}} {\rm d} {r} \int \int {\rm d} \zeta {\rm d} \bar{\zeta}
{R^2} \left( R' - R \frac{{\cal E}'}{{\cal E}} \right) \frac{1}{{\cal E}^2}  \nonumber \\
&& = 
\int\limits_{r_c}^{r_{\mathcal{D}}} {\rm d} {r}
\left[ {R^2 R'} 
\int \int \frac{{\rm d} \zeta {\rm d} \bar{\zeta}}{{\cal E}^2} +
\frac{1}{2} {R^3 } \frac{\partial}{\partial  {r}} \left( 
 \int \int \frac{{\rm d} \zeta {\rm d} \bar{\zeta}}{{\cal E}^2} \right) \right].
\end{eqnarray}
where $r_c$ is the lower limit of integration (only for the quasi--spherical
and quasi--planar model $r_c = 0$),  
${\rm d} \zeta {\rm d} \bar{\zeta}/{\cal E}^2$ is the metric of a unit sphere/plane/hyperboloid and
\[ \int \int \frac{{\rm d} \zeta {\rm d} \bar{\zeta}}{{\cal E}^2} = S_\mathcal{D}, \]
which does not depend on $r$ (for the quasi--spherical model $S_\mathcal{D} = 4\pi$). Thus 
\begin{equation}
V_{q} = S_\mathcal{D} \int\limits_{r_c}^{r_{\mathcal{D}}} {\rm d}  {r} {R^2 R'} \equiv S_\mathcal{D} \Upsilon_\mathcal{D}
\label{vol} 
\end{equation}
Note that $R(r_c)$ can be equal zero \cite{HK08}, so
$ \Upsilon_\mathcal{D} = (1/3) R_\mathcal{D}^3$, even if for the 
quasi--hyperbolic and quasi--planar model $r$ cannot be equal to $0$.
The q--density  $\rho_q$ is
\begin{eqnarray}
&& \rho_q  = \frac{1}{V_{q}} \int\limits_{r_c}^{r_{\mathcal{D}}} {\rm d}  {r} \int \int {\rm d} \zeta {\rm d} \bar{\zeta} ~
\EE Y' Y^2 \rho  \nonumber \\
&& = \frac{1}{S_\mathcal{D} \Upsilon_\mathcal{D}} \int\limits_{r_c}^{r_{\mathcal{D}}} {\rm d}  {r} \int \int \frac{{\rm d} \zeta {\rm d} \bar{\zeta}}{{\cal E}^2} {R^2}
\left( R' - R \frac{{\cal E}'}{{\cal E}} \right)
\frac{2M' - 6 M {\cal E}'/{\cal E}}{R^2 \left( R' - R {\cal E}'/{\cal E} \right)}  \nonumber \\
&& = \frac{2}{\Upsilon_{\mathcal{D}}} \int\limits_{r_c}^{r_{\mathcal{D}}} {\rm d}  {r} {M'}  + 
\frac{1}{3 S_\mathcal{D} \Upsilon_\mathcal{D}} \int\limits_{r_c}^{r_{\mathcal{D}}} {\rm d}  {r} {M}
 \frac{\partial}{\partial  {r}} \left( 
 \int \int \frac{{\rm d} \zeta {\rm d} \bar{\zeta}}{{\cal E}^2} \right)  \nonumber \\
&& = \frac{1}{\Upsilon_{\mathcal{D}}} \int\limits_{r_c}^{r_{\mathcal{D}}}{\rm d}  {r} {M'} = \frac{3 M_\mathcal{D}}{R_\mathcal{D}^3}
\label{arho}
\end{eqnarray}
\noindent The q--scalar $\Theta_q$ dual to the local expansion $\Theta$ is

\begin{eqnarray}
&& \Theta_q = \frac{1}{V_{q}} \int\limits_{r_c}^{r_{\mathcal{D}}}{\rm d}  {r} \int \int \frac{ {\rm d} \zeta {\rm d} \bar{\zeta}}{{\cal E}^2}
{R^2} \left( R' - R \frac{{\cal E}'}{{\cal E}} \right)  \frac{\dot{R'} + 2 \dot{R}{R'}/ R - 3 \dot{R} {\cal E}'/{\cal E}}{R' - R {\cal E}'/{\cal E}}   \nonumber \\ 
&& = \frac{1}{\Upsilon_{\mathcal{D}}} \int\limits_{r_c}^{r_{\mathcal{D}}}{\rm d}  {r} 
 {R^2 R'} \left(
 \frac{\dot{R}'}{R'} + 2 \frac{\dot{R}}{R} \right) = 
\frac{\dot{R}_\mathcal{D}}{R_\mathcal{D}}
\label{atht}
\end{eqnarray}

\noindent Finally, the q--scalar $\RR_q$ dual to $\RR$ is 

\begin{eqnarray}
&& ^3\mathcal{R}_q =  \frac{1}{V_q} \int\limits_{r_c}^{r_{\mathcal{D}}}{\rm d}  {r} \int \int \frac{ {\rm d} \zeta {\rm d} \bar{\zeta}}{{\cal E}^2}
 \left( R' - R \frac{{\cal E}'}{{\cal E}} \right)  {2 K} 
  \left( \frac{ R K'/K - 2 R  {\cal E}'/{\cal E}}{
R' - R {\cal E}'/{\cal E}} + 1 \right)   \nonumber \\ 
 && =  \frac{2 }{\Upsilon_{\mathcal{D}}} \int\limits_{r_c}^{r_{\mathcal{D}}}{\rm d}  {r} {( R K )'} = 6 \frac{K_{\mathcal{D}}}{R_{\mathcal{D}}^2} .
\label{a3r}
\end{eqnarray}
Note that (\ref{arho}), (\ref{atht}) and (\ref{a3r}) exactly coincide with (\ref{rhoq}), (\ref{Thetaq}) and (\ref{RRq}) at the domain boundary marked by $r$.

\section{Covariant meaning of the q--scalars and their fluctuations.}\label{covariant}

 It is straightforward to show that the q--scalars $\rho_q,\,\HH_q,\,\KK_q,\,\Omega_q$ and their fluctuations are coordinate independent quantities expressible in terms of curvature invariants. The q--scalar $\rho_q$ and the fluctuation $\Dm$ take the form
\begin{equation} 8\pi\rho_q = 6\Psi_2-{\cal{R}},\qquad \Dm = \frac{6\Psi_2/{\cal{R}}}{1-6\Psi_2/{\cal{R}}},\label{rhoDrho}\end{equation}
where ${\cal{R}}=-8\pi\rho$ is the 4--dimensional Ricci scalar and $\Psi_2=-\W$ is the only nonzero Weyl curvature invariant in a Newman--Penrose tetrad. The fact that the ratio of Weyl to Riemann curvature scalars can provide a measure of inhomogeneity by means of suitably defined density fluctuations in LTB models has been already highlighted  in \cite{wainwright}. Expressions similar to (\ref{rhoDrho}) follow from (\ref{locvars2}) for $\HH_q$ and $\Dh$:
\begin{equation} \HH_q = \HH +\Sigma,\qquad \Dh = -\frac{\Sigma/\HH}{1+\Sigma/\HH},\label{HDH}\end{equation}
where  $\HH=3\Theta=h_a^b\nabla_bu^a$ is the expansion scalar and $\Sigma$  is the scalar in (\ref{Sigma1}) associated with the shear tensor. While $\Sigma$ is already given in a coordinate independent form in (\ref{Sigma1}), it is easier to provide an invariant characterization of it as the independent eigenvalue of the shear tensor $\sigma^a\,_b$. Notice that the form of $\Dh$ above is directly related to the scalar ratio $\Sigma/\Theta$ or the quadratic form $\sigma_{ab}\sigma^{ab}/\Theta^2=6\Sigma^2/\Theta^2$ which provide  coordinate independent measures of the ratio of anisotropic vs isotropic local expansion velocities.   Coordinate independent forms for the remaining q--scalars $\KK_q,\,\Omega_q$ and fluctuations $\Dk,\,\DOm$ follow by applying the constraints (\ref{FFqham})--(\ref{FFqcon}) and (\ref{Omdef})--(\ref{DOmdef}) to (\ref{rhoDrho}) and (\ref{HDH}).

\section{Functions of q--scalars are q--scalars.}\label{qfunctions}

Let $\psi = \psi(A_q,B_q)$  be a smooth function of $A_q$ and $B_q$ complying with (\ref{qdef}), then considering (\ref{Aqr}) and (\ref{Dadef}), \, $\psi=\phi_q$ is the q--scalar for which we identify $\Delta^{(\phi)}$ and $\phi$ by:
\ba \fl\Delta^{(\phi)} =\frac{A_q}{\psi}\frac{\dd\psi}{\dd A_q}\Delta^{(A)}+\frac{B_q}{\psi}\frac{\dd\psi}{\dd B_q}\Delta^{(B)},\nonumber\\
\fl\phi = \phi_q\,[1+\Delta^{(\phi)}]=\psi+A_q\frac{\dd\psi}{\dd A_q}\Delta^{(A)}+B_q\frac{\dd\psi}{\dd B_q}\Delta^{(B)}=\psi+\frac{\psi'}{3Y'/Y}.\nonumber\ea         
This property was used to define $\DOm$ and $\Omega$ in (\ref{DOmdef}), since $\Omega_q=\Omega_q(\rho_q,\HH_q)$ in (\ref{Omdef}). However, as opposed to the local scalars $\rho,\,\HH,\,\KK$, the physical and geometric interpretation of the ``local'' scalar $\Omega$ is not clear.

\section{Models with ``closed'' and ``wormhole'' topologies.}\label{wormholes}

The coordinate choice (\ref{Ridef}) cannot be used if the slices $\T[t]$ are ``closed'' (homeomorphic to $\mathbb{S}^3$) or are ``wormholes'' (homeomorphic to either $\mathbb{S}^2\times \mathbb{R}$ or $\mathbb{S}^2\times \mathbb{S}^1$). In the former case there are two worldlines where $R_0=R=0$, which generalize the closed spherical models with two symmetry centers and in the latter $R_0$ has no zeroes. In both cases $R'_0$ must have a zero. Hence, we choose:
\begin{equation}
R_0 = \ell_0\, f(r),\label{Ridef2}
\end{equation}
where $\ell_0$ is an arbitrary length scale and $f(r)$ is a dimensionless function with the appropriate properties. For closed models $f$ can be a sine--type of function, while for the wormhole case it can be either $\cosh$ or $\sec$. The coordinate choice (\ref{Ridef2} implies the following replacement
\begin{equation}
\frac{r\EE'}{\EE} \quad \to \quad \frac{\EE'/\EE}{f'/f} = \frac{\dd\ln\EE}{\dd\ln f}
\end{equation}
hence, $f'$ and $\EE'$ (and thus $S',\,P',\,Q'$) must have same order common zeroes.

\end{appendix}

\section*{Acknowledgements}
We acknowledge financial support from grant PAPIIT--DGAPA IN-119309. 
We also acknowledge the support from the European Union Seventh Framework Programme under 
the Marie Curie Fellowship (PIEF-GA-2009-252950).


\section*{References}
      
 \begin{thebibliography}{00}
 


\bibitem{Step2003}
Stephani H Kramer D MacCallum MAH Hoenselaers C and
Herlt E {\it Exact solutions of Einstein's field equations}, 2nd edn.
Cambridge University Press, Cambridge (2003)

\bibitem{Lema1933}
Lema\^{\i}tre G 1933 {\it Ann. Soc. Sci. Bruxelles} A {\bf 53} 51 (English
translation, with historical comments: 1997 {\it Gen. Rel. Grav.} {\bf 29} 637)

\bibitem{Tolm1934}
Tolman R C 1934 {\it Proc. Nat. Acad. Sci. USA} {\bf 20} 169 (reprinted, with
historical comments: 1997 {\it Gen. Rel. Grav.} {\bf 29} 931)

\bibitem{Bond1947}
Bondi H 1947 {\it Mon. Not. R. Astron. Soc}. {\bf 107} 410 (reprinted with
historical introduction in 1999 {\it Gen. Rel. Grav.} {\bf 11} 1783)

\bibitem{BKHC2009} Bolejko K Krasi\'nski A Hellaby C and C\'el\'erier M N 2009 {\it Structures in the Universe by exact methods: formation, evolution, interactions} Cambridge University Press, Cambridge

\bibitem{BCK2011}
Bolejko K C\'el\'erier M N and Krasi\'nski A 2011
{\it Class. Quantum Grav.} {\bf 28} (2011) 164002.


\bibitem{S75} Szekeres P 1975 {\it  Commun Math Phys} {\bf 41} 55.

\bibitem{Sz75} Szekeres P 1975 {\it Phys Rev} D {\bf 12} 2941

\bibitem{barrow1} Barrow J D B and Stein--Schabes J 1984 {\it Phys Lett} A {\bf 103} 315-318

\bibitem{barrow2} Barrow J D B and Saich P
1993 {\it Mon Not Roy Astr Soc} {\bf 262} 717-725

\bibitem{GoWa1982}
Goode S W and  Wainwright J 1982 {\it Phys Rev} D {\bf 26} 3315


\bibitem{HK02} Hellaby C and Krasi\'nski A 2002 {\it Phys Rev} D {\bf 66} 084011

\bibitem{Bsz1} Bolejko K 2006 {\it Phys Rev} D {\bf 73} 123508.

\bibitem{K08}  Krasi\'nski A 2008 {\it Phys Rev} D {\bf 78} 064038.

\bibitem{HK08} Hellaby C and  Krasi\'nski A 2008 {\it Phys Rev} D {\bf 77} 023529


\bibitem{PK06} Pleba\'nski J and Krasi\'nski A 2006  {\it An Introduction to General Relativity and Cosmology}. Cambridge University Press, Cambridge.

\bibitem{IRGW2008}
Ishak M Richardson J Garred D Whittington D
Nwankwo A and Sussman R A 2008 {\it Phys Rev} D {\bf 78} 123531

\bibitem{Bole2009-cmb}
Bolejko K 2009 {\it Gen Rel Gravit} {\bf 41} 1737

\bibitem{BoCe2010}
Bolejko K and C\'el\'erier M N 2010 {\it Phys Rev} D {\bf 82} 103510

\bibitem{NIT2011}
Nwankwo A, Ishak M, and Thompson J, 2011, {\it J. Cosmol. Astropart. Phys.} 1105, 028

\bibitem{NMMB2011}
Meures N and Bruni M 2011, arXiv:1107.4433

\bibitem{Bsz2} Bolejko K 2007 {\it Phys Rev} D {\bf 75} 043508.


\bibitem{IsPe2011}
Ishak M and Peel A 2011 arXiv:1104.2590

\bibitem{MeBr2011}
Meures N and Bruni M 2011 {\it Phys Rev} D {\bf 83}  123519

\bibitem{KrBo2010}
Krasi\'nski A and  Bolejko K {\it Phys Rev} D {\bf 83} 083503

\bibitem{BoSu2011}
Bolejko K and Sussman R A 2011 {\it Phys Lett} B {\bf 697} 265


\bibitem{Szafron}
Szafron D A 1977 {\it J. Math. Phys}. {\bf 18}, 1673

\bibitem{kras97} Krasi\'nski A 1997 {\it  Inhomogeneous Cosmological Models, Cambridge University Press}, Cambridge University Press, Cambridge.


\bibitem{suss09} Sussman R A 2009 {\it Phys Rev} D {\bf 79} 025009


\bibitem{lw1} Deruelle N and Langlois D 1995 {\it Physical Review} D {\bf 52} 2007

\bibitem{lw2} Sopuerta C F {\it et al} 1999 {\it Phys Rev} D {\bf 60} 24006.

\bibitem{lw3} Bruni M and Sopuerta C F 2003 {\it Class Quantum Grav} {\bf 20} 5275--5290.


\bibitem{silent1} van Ellst H {\it et al} 1997 {\it Class Quant Grav} {\bf 14} 1151-1162

\bibitem{silent2} Sopuerta C F 1997 {\it Phys Rev} D {\bf 55} 5936


\bibitem{suss2009} Sussman R A 2009 {\it Phys Rev} D {\bf 79} 025009.

\bibitem{sussIU} Sussman R A 2010 {\it AIP Conf Proc} {\bf 1241} 1146-1155 ({\it Preprint} {\tt arXiv:0912.4074 [gr-qc]})

\bibitem{suss2010a} Sussman R A 2010 A new approach for doing theoretical and numeric work with Lema\^{\i}tre--Tolman--Bondi dust models ({\it Preprint} {\tt arXiv:1001.0904v1})


\bibitem{LTBave1} Paranjape A and Singh T P 2006 {\it Class.Quant.Grav.},{\bf 23}, 6955ÔøΩ6969

\bibitem{LTBave2} Chuang CH, Gu J A and Hwang W Y P 2008 {\it{Class. Quant. Grav.}} \textbf{25} 175001

\bibitem{sussBR1} Sussman R A 2008 On spatial volume averaging in Lema\^itre--Tolman--Bondi dust models. Part I: back reaction, spacial curvature and binding energy  ({\it Preprint} {\tt arXiv:0807.1145 [gr-qc]})

\bibitem{sussBR2} Sussman R A 2011 
{\textit{Class.Quant.Grav.}} \textbf{28} 235002 ({\it Preprint} {\tt arXiv:1102.2663 [gr-qc]})



\bibitem{RadAs} Sussman R A 2010 {\textit{Gen Rel Grav}} {\bf{42}} 2813--2864 ({\it Preprint} {\tt arXiv:1002.0173 [gr-qc]})

\bibitem{RadProfs} Sussman R A 2010 {\textit{Class.Quant.Grav.}} \textbf{27} 175001 ({\it Preprint} {\tt arXiv:1005.0717 [gr-qc]})


\bibitem{ds1} Sussman R A 2008 {\it Class Quant Grav} {\bf 25} 015012  ({\it Preprint} {\tt arXiv:0709.1005})

\bibitem{ds2} Sussman R A 2011 {\it Class Quant Grav} {\bf 28} 045006  ({\it Preprint} {\tt arXiv:1004.0773})


\bibitem{baro} Barnes A and Rowlingson R R 1989 {\it Class Quantum Grav} {\bf 6} 949--960.


\bibitem{MPS1} Matarrese S Pantano O and Saez D 1993 {\it Phys Rev} D {\bf 47}

\bibitem{MPS2} Matarrese S Pantano O and Saez D 1994 {\it Phys Rev Lett} {\bf 72} 320–323.

\bibitem{bruni1} Bruni M Matarrese S and Pantano O 1995 {\it Astroph Jou} {\bf 445} 958

\bibitem{bruni2} Bruni M Matarrese S and Pantano O 1995 {\it Phys Rev Lett} {\bf 74} 1916.

\bibitem{croud} Croudace K M {\it et al} 1994 {\it Astroph Jou} {\bf 423} 22–32.


\bibitem{1plus3} Ellis G F R and Bruni M 1989 {\it Phys. Rev.} D {\bf 40}  1804;  Ellis G F R and van Elst H 1998 Cosmological Models (Carg\`ese Lectures 1998) ({\it Preprint} {\tt arXiv gr-qc/9812046 v4})

\bibitem{zibin}  Zibin J P 2008 {\it Phys Rev} D{\bf 78} 043504 [arXiv:0804.1787]

\bibitem{dunsbyetal} Dunsby P {\it et al} ({\it Preprint} {\tt  arXiv:1002.2397v1 [astro-ph CO]})



\bibitem{hayward} Hayward S A 1996 {\it Phys Rev} D {\bf 53} 1938 ({\it Preprint} {\tt ArXiv gr-qc/9408002}); Hayward S A 1998 {\it Class Quantum Grav} {\bf 15} 3147ÔøΩ3162  ({\it Preprint} {\tt ArXiv gr-qc/9710089v2})


\bibitem{Bole2009-vav}
 Bolejko K 2009 {\it Gen Rel Gravit} {\bf 41} 1585


\bibitem{moffat} Moffat J W 2006 {\it J. Cosmol. Astropart. Phys.} JCAP(2006)001;

\bibitem{kolb} Kolb E W, Matarrese S, Notari A and  Riotto A 2005 {\it Phys Rev} D {\bf 71} 023524  ({\it Preprint} {\tt arXiv:hep-ph/0409038v2}); Marra V,  Kolb E W and Matarrese S 2008 {\it Phys Rev} D {\bf 77} 023003; Marra V, Kolb E W, Matarrese S and Riotto A 2007 {\it Phys Rev} D {\bf 76} 123004.

\bibitem{alnes} Alnes H, Amazguioui M and Gron O 2006 {\it Phys Rev} D {\bf 73} 083519; Alnes H and Amazguioui M 2006 {\it Phys Rev} D {\bf 74} 103520;  Alnes H and Amazguioui M 2006 {\it Phys Rev} D {\bf 75} 023506

\bibitem{endqvist} Enqvist K and  Mattsson T 2007 {\it JCAP} {\bf 0702} 019 ({\it Preprint} {\tt        arXiv:astro-ph/0609120v4}); Enqvist K 2008 {\it Gen. Rel. Grav.}  {\bf 40}  451-466 ({\it Preprint} {\tt arXiv:0709.2044})

\bibitem{kbsn1}
Bolejko K 2008 {\it PMC Physics} A2, 1

\bibitem{GBH} Garc\'\i a--Bellido J and Troels H 2008 {\bf JCAP} 0804:003 ({\it Preprint} {\tt gr-qc/0802.1523v3 [astro-ph]})

\bibitem{kbsn2}
Bolejko K and Wyithe J S B 2009 {\bf JCAP} 0902:020


\bibitem{clarkson} February S {\it et al}  et al (dark energy void) ({\it Preprint} {\tt arXiv:0909.1479v2[astro-ph CO]})


\bibitem{ltbstuff1} Hellaby C and Lake K 1985 {\it Astrophys J.} {\bf 290} 381

\bibitem{ltbstuff2} Matravers D R and Humphreys N P  2001 {\it Gen. Rel. Grav.} {\bf 33}
531-52; Humphreys N P, Maartens R and Matravers D R 1998 Regular spherical dust spacetimes {\tt Preprint gr-qc/9804023v1}.


\bibitem{wainwright} Wainwright J and Andrews S 2009 {\it Class.Quant.Grav.},{\bf 26}, 085017


 \end{thebibliography}
\end{document}